\documentclass[prd,nofootinbib,superscriptaddress,showpacs]{revtex4}

\usepackage{amsmath,amssymb}
\usepackage{graphicx}
\usepackage{dcolumn}
\usepackage{bm}
\def\be{\begin{eqnarray}}
\def\ee{\end{eqnarray}}
\newcommand{\bea}{\begin{eqnarray}}
\newcommand{\eea}{\end{eqnarray}}
\newcommand{\nn}{\nonumber}
\newcommand{\dbyd}[1]{\ensuremath{\frac{d\phantom{#1}}{d #1}}}

\def\d{{\rm d}}
\def\ie{{\it i.e.}}

\begin{document}
\title{Self--accelerating Warped Braneworlds}
\date{\today}

\preprint{FERMILAB-PUB-06-422-T}
\preprint{DRAFT V5, \today}

\author{Marcela Carena}
\affiliation{ Fermi National Accelerator
Laboratory, P.O. Box 500,  Batavia, IL 60510, USA}
\author{Joseph Lykken}
\affiliation{ Fermi National Accelerator
Laboratory, P.O. Box 500,  Batavia, IL 60510, USA}
\author{Minjoon Park}
\affiliation{
Department of Physics, University of California, Davis,
CA 95616, USA
}
\author{Jos\'e Santiago}
\affiliation{ Fermi National Accelerator
Laboratory, P.O. Box 500,  Batavia, IL 60510, USA}

\begin{abstract}
Braneworld models with induced gravity have the potential to
replace dark energy as the explanation for the current
accelerating expansion of the Universe. The original model
of Dvali, Gabadadze and Porrati (DGP) demonstrated the existence
of a ``self--accelerating'' branch of background solutions,
but suffered from the presence of ghosts. We present a new large class of 
braneworld models which generalize the DGP model.
Our models have negative curvature in the bulk, allow a second brane, and
have general brane tensions and localized curvature terms. 
We exhibit three different kinds of ghosts, associated to
the graviton zero mode, the radion, and the longitudinal components
of massive graviton modes. The latter two species occur in the
DGP model, for negative and positive brane tension respectively. 
In our models, we find that the two kinds of DGP ghosts are
tightly correlated with each other, but are not always linked
to the feature of self--acceleration. Our models
are a promising laboratory for understanding the origins and
physical meaning of braneworld ghosts, and perhaps for eliminating
them altogether.

\end{abstract}


\maketitle

\section{Introduction}

Models with extra spatial dimensions provide new perspectives on
many long standing problems of particle physics and gravity.
One of the most exciting suggestions
\cite{Deffayet:2000uy,Deffayet:2001pu}  
is that an extra
dimension, rather than dark energy, may be the origin of 
the currently accelerating expansion of
the Universe. The original simple model of Dvali, Gabadadze and Porrati
(DGP) posits that we live on a codimension one brane in an
infinite flat five-dimensional bulk \cite{Dvali:2000hr}. As
generalized in \cite{Koyama:2005tx},
the model has only two relevant input parameters: a crossover scale
$r_c$ determined by the ratio of the 4d brane-localized gravity
coupling with the 5d bulk one, and the tension of the brane.
The Einstein equations have two kinds of solutions: a normal
branch and a ``self--accelerating'' branch; both branches cause
a modification of the effective 4d Friedmann equation, with the second
one having the property that at sufficiently late times it gives
an accelerated expansion without the need of dark energy to drive it.

The DGP model has encountered many difficulties, but the one that
looks most serious is the presence of ghosts in the weakly--coupled 
long distance regime of the theory linearized
around the self--accelerating background 
solution \cite{Koyama:2005tx}-\cite{Nicolis:2004qq}.
The physical origin and meaning of these ghosts are not
yet clear. 

Ghosts, like tachyons, are often an indication of an instability
of the theory perturbed around a given background. Indeed it
is not surprising that gravitating systems with nontrivial backgrounds
including branes (in the absence of supersymmetry) exhibit a variety
of instabilities. In the case of DGP, it is natural to ask whether
modified brane setups might avoid ghosts and tachyons, while
still exhibiting the key feature of self--acceleration.

In this paper we present a large class of warped brane setups 
that generalize the DGP model. Our basic idea is to use an
$AdS_5$ bulk space, taking advantage of the fact that slices
of $AdS_5$ can be flat, $AdS_4$, or $dS_4$. We consider models
with one or two branes, taking the brane tensions and the
brane-localized gravity couplings as input parameters, along
with the bulk cosmological constant. We obtain background solutions
with both normal and self--accelerating branches, for branes that
are either $AdS_4$ or $dS_4$. The self--accelerating branches for
the de Sitter branes give models precisely analogous to the DGP
model; the self--accelerating branches for $AdS_4$ branes are
not of obvious cosmological interest, but allow us to generalize
the well-understood physics of the Karch-Randall model \cite{Karch:2000ct}.

In two previous papers \cite{Carena:2005gq,Bao:2005bv}
we set up the necessary tools to study 
generic warped models
with arbitrary brane
tensions and localized curvature terms.
The size of the extra dimension and the
brane curvature are well defined functions of the input
brane parameters,
and therefore by varying the
latter we can parametrize a large number of different models.
In one limit we make explicit contact with the original DGP model,
while in another limit we reproduce the $AdS_5/AdS_4$ Karch-Randall model. 
In \cite{Carena:2005gq,Bao:2005bv} we studied the phenomenology
of $AdS_4$ branes for background solutions on the standard branch,
exhibiting a variety of ghosts and tachyons in parts of the parameter
space of models. 
In this article we will
extend that study to the self--accelerating branch and also consider
the case in which the branes are $dS_4$.
We will pay particular attention to the presence of ghosts in the
spectrum and how this is related to which branch we are considering.

The outline of the paper is the following. In Section~\ref{ghosts}
we describe some basic facts about ghosts in theories with gravity. 
In Section~\ref{model} we
review the main features of our new models, extending the results of
\cite{Carena:2005gq,Bao:2005bv} to the self--accelerating branch
and $dS_4$ branes. The spectrum, with particular emphasis on the
presence of ghosts and tachyons, is studied for the different cases in
Section~\ref{spectrum}. Section~\ref{conclusions} is devoted to our
conclusions, and some technical details are relegated to appendices.

\section{Ghosts and Gravity\label{ghosts}}

With the $\mbox{({--}+++)}$ metric signature, a canonical scalar
field kinetic action in flat space is written
\be
\int d^4x\,\left( -\frac{1}{2}\eta^{\mu\nu}\partial_\mu \phi
\partial_\nu \phi - \frac{1}{2}m^2 \phi^2 \right) \; .
\ee
Reversing the sign of the first term produces a ghost:
\be\label{eqn:sglag}
\int d^4x\,\left( \frac{1}{2}\eta^{\mu\nu}\partial_\mu \phi
\partial_\nu \phi - \frac{1}{2}m^2 \phi^2 \right) \; .
\ee
As discussed \textit{e.g.} in \cite{Cline:2003gs}, defining
the ghost propagator with the usual Feynman $i\epsilon$
prescription results in a nonunitary theory with negative norm states;
defining the propagator with the opposite $i\epsilon$ prescription
preserves unitarity but induces a catastrophic instability from
propagating negative energy states.

Thus kinetic ghosts are similar to tachyons, in that they can be
regarded as instabilities. Indeed some ghosts are also tachyons:
this is the case in (\ref{eqn:sglag}) for $m^2>0$, since 
the solutions of the wave equation will be exponentials rather
than plane waves.
As with tachyons,
in some cases it may be possible to cure a ghost instability by
perturbing around a different ground state. Such a shift
can be thought of as the nonperturbative formation
of a tachyon or ghost condensate; a tachyon condensate may  produce
a stable static ground state while a ghost condensate may produce
a time dependent but ghost--free ground state \cite{Arkani-Hamed:2003uy}.

In addition to ghosts and tachyons, one might worry that simple
scalar field theories can suffer from other diseases. Suppose we
add a derivative self-interaction term to (\ref{eqn:sglag}):
\be\label{eqn:sgl}
\int d^4x\,\left( -\frac{1}{2}\eta^{\mu\nu}\partial_{\mu} \phi
\partial_{\nu} \phi - \frac{1}{2}m^2 \phi^2 
+ \frac{c}{2\Lambda^4}(\eta^{\mu\nu}\partial_{\mu} \phi
\partial_{\nu} \phi)^2
\right) \; ,
\ee
where $\Lambda$ is the UV cutoff of this effective theory.
For $c < 0$, this theory appears to have analyticity problems
and exhibits superluminal modes in the massless limit, 
despite a lack of tachyonic 
instability \cite{Adams:2006sv}. However
Jenkins and O'Connell have recently shown \cite{Jenkins:2006ia} that 
``positivity constraints'' of this type are equivalent to 
a no--tachyon condition in a (partial) UV completion. In other
words, effective theories of the form (\ref{eqn:sgl}) with
$c < 0$ arise from integrating out tachyons before curing
the tachyonic instability. Thus
kinetic ghost and tachyon instabilities appear to exhaust the
physically distinct diseases of scalar field theories.

This simple picture, however, is intrinsic to flat space propagators
and flat space kinetic terms. Adding gravity to the picture
complicates the discussion of ghosts, and has led to much
debate in the literature.
Once we add gravity, we must be on the lookout for new kinds of
ghosts. In addition to scalar ghosts, the longitudinal mode of 
a massive graviton can be a ghost. As we will see, even a
massless graviton can be a ghost. Because gravity is a nonlinear theory
with a large local symmetry, we must also work harder to
understand the physical significance of ghosts. 

\subsection{de Sitter ghosts}

It was shown many years ago by Higuchi \cite{Higuchi:1986py} that the
longitudinal modes of massive gravitons can be ghosts if we
are in a de Sitter background. Letting $12\mathcal{H}^2$ denote the constant
de Sitter curvature, massive gravitons with mass in the
range $0 < m^2 < 2\mathcal{H}^2$ are ghosts. These correspond
to nonunitary representations of the de Sitter group $SO(4,1)$,
as discussed in \cite{Higuchi:1986wu,Garidi:2003bg}, but
the analysis of Deser and Waldron \cite{Deser:2001wx} shows
that they are also kinetic ghosts. As we will see in the next
section using an explicit on-shell tensor decomposition, the
effective kinetic term of a longitudinal massive graviton
mode $s(x)$ in a $dS_4$ background is given by:
\be\label{eqn:lmodekin}
\frac{3}{2} m^2(m^2-2\mathcal{H}^2)\int d^4x\; s(x)\left( -\partial_0^2 +
\nabla_i^2 - ( m^2 -\frac{9}{4}\mathcal{H}^2)
\right)s(x)
\; ,
\ee
where $\nabla_i^2$ is the de Sitter Laplacian
$\nabla_i^2 = \partial_i^2/f(t)^2$, with
$f(t) = $ exp$(\mathcal{H}t)$ coming from the
$dS_4$ background metric: $g_{00} = -1$, $g_{ij} = f^2\delta_{ij}$.
From (\ref{eqn:lmodekin}) we see immediately that the
longitudinal mode is a ghost for $0 < m^2 < 2\mathcal{H}^2$.
It would also naively appear that the
longitudinal mode is a tachyon for $0 < m^2 < \frac{9}{4}\mathcal{H}^2$.
However the analysis of tachyonic instabilities in $dS_4$ is
complicated by the fact that the naive Hamiltonian corresponding
to the wave operator in (\ref{eqn:lmodekin}) has an explicit 
time-dependence from $f(t)$ and is therefore not conserved.
The operator corresponding to the actual conserved energy is
manifestly positive \textit{inside the de Sitter horizon} for any $m^2 > 0$,
as shown in \cite{Deser:2001wx}. This is referred to as a
``mild'' tachyonic instability in the last reference in
\cite{Koyama:2005tx};
for our purposes we will not call such modes tachyons, reserving
that name for cases such as $m^2 < 0$ in (\ref{eqn:lmodekin})
which resemble tachyons in flat space.

It was shown in \cite{Koyama:2005tx}
that the DGP
model with positive brane tension contains 
a longitudinal massive graviton mode ghost on the self--accelerating
branch. 
Intriguingly, for negative brane tension
the massive graviton ghost is absent, replaced instead
by a ghost radion scalar. In the warped models
presented in the paper, we will find that these two kinds of ghosts
are also tightly related. 

The physical interpretation of ghosts in theories with gravity
can be obscured by the presence of local symmetries.
For example the DGP model in the limit of a tensionless brane
has a massive graviton mode with $m^2=2H^2$; the resulting
4d massive gravity theory has an enhanced ``accidental'' local symmetry
beyond 4d general covariance \cite{Deser:1983mm,Deser:2001pe}.
Charmousis \textit{et al.} 
found a ghost in this limit
as well \cite{Koyama:2005tx}, 
but Deffayet \textit{et al} have argued \cite{Deffayet:2006wp}
that this mode
is actually a Lagrange multiplier enforcing an explicit
gauge-fixing of the extra symmetry. This is an interesting 
special case to be considered elsewhere.

\subsection{when a ghost is not a ghost}

It can be argued (incorrectly) that gravity renders
the physical meaning of ghosts ambiguous.
Consider, for example, pure 4d Einstein gravity.
Perform a conformal rescaling of the metric by the
field redefinition 
\be\label{eqn:shift}
g_{\mu\nu} \to \left(1+\frac{\phi}{M} \right)g_{\mu\nu} \; ,
\ee
where $\phi (x)$ is a 4d scalar field and $M$ is the reduced Planck mass.
Now substitute the rescaled metric into the Einstein-Hilbert
action, assuming that the background metric before
rescaling was flat:
\be
\sqrt{-g}\,M^2R
\to 
\frac{3}{2}\,\frac{1}{1+\phi/M}\,\eta^{\mu\nu}\partial_{\mu}\phi
\partial_{\nu}\phi
\; .
\ee
This looks like a kinetic ghost. Now suppose we start with a Lagrangian
whose perturbation theory around flat space is obviously ghost-free:
\be\label{eqn:gpluss}
\int d^4x\, \sqrt{-g}\, \left( M^2R -
\frac{1}{2}\,g^{\mu\nu}\partial_{\mu}\phi
\partial_{\nu}\phi \right)
\; .
\ee
It appears naively that by a simple field redefinition of the metric
we can produce a ghost.

The flaw in this argument is that we have ignored the subtleties
of general covariance and gauge-fixing \cite{Deffayet:2006wp}. 
Consider again
4d Einstein gravity expanded around flat space:
$g_{\mu\nu} = \eta_{\mu\nu} + h_{\mu\nu}$.
As discussed in \cite{Carena:2005gq}, the massless symmetric
tensor $h_{\mu\nu}$ can be decomposed as
\be
h_{\mu\nu} = \beta_{\mu\nu} + \partial_{\mu}v_{\nu}+
\partial_{\nu}v_{\mu} +\partial_{\mu}\partial_{\nu}\varphi_1
+c_{\mu\nu} + \partial_{\mu}n_{\nu} +\partial_{\nu}n_{\mu}
+\eta_{\mu\nu}\varphi_2 \; .
\ee
Here $\beta_{\mu\nu}$ is traceless, transverse, and orthogonal
to $n_{\mu}$, giving 2 degrees of freedom. Similarly $v_{\mu}$
is transverse and orthogonal to $n_{\mu}$, giving 2 degrees of
freedom, while the longitudinal null vector $n_{\mu}$  
gives 1 more degree of freedom. Lastly, $c_{\mu\nu}$ is traceless 
but not transverse, giving 3 degrees of freedom, which together with
the two scalars $\varphi_1$ and $\varphi_2$ adds up to the total 10
degrees of freedom of $h_{\mu \nu}$.

Obviously $v_{\mu}$, $n_{\mu}$ and $\varphi_1$ are the 4 pure gauge modes of
4d general covariance, while $\beta_{\mu\nu}$ gives the two traceless
transverse 
propagating degrees of freedom of an on-shell massless graviton.
The four remaining degrees of freedom represented by $\varphi_2$ and
$c_{\mu\nu}$ are subject to constraints from the equations of motion.
In the absence of sources they are constrained to vanish.
Just as for the more familiar case of the time-like component of an 
abelian gauge field, these modes do not really propagate. 
In the post-Newtonian approximation \cite{wbook},
$\varphi_2$ is proportional
to the Newtonian scalar potential (analogous to the electrostatic potential)
while $c_{\mu\nu}$ contains the
Newtonian vector potential.
So $\varphi_2$ represents the
same kind of ghost as the time-like component of an abelian
gauge field (which in the nonrelativistic limit gives the electrostatic
potential). Such ghosts are not really ghosts since they do not
propagate: the corresponding degrees of freedom are completely
fixed in terms of sources by solving constraints from the equations
of motion.

Thus the metric rescaling (\ref{eqn:shift}) in the theory
described by (\ref{eqn:gpluss}) mixes a non-ghost matter scalar
with a pseudo-ghost mode of the metric. The full quadratic
action of the rescaled theory will have kinetic terms mixing
$\phi$ with $\varphi_2$. Presented with such an action
\textit{ab initio}, the proper way to extract its physical degrees
of freedom is to gauge--fix it and diagonalize the kinetic terms.
The only way to diagonalize the kinetic terms is
by shifting $\varphi_2 \to \varphi_2 - \phi/M$, 
effectively undoing the metric rescaling (\ref{eqn:shift}). 
The theory is then obviously free of ghosts.

\subsection{summary}

The moral of this story is that there is an unambiguous procedure for
determining the presence of ghosts in a theory with gravity expanded
around a given background solution.
First, determine the local symmetries and gauge--fix them.
Second, diagonalize the full quadratic action.
Third, extract the propagators of the physical degrees of freedom 
and check for kinetic ghosts.  
It is also important to check for tachyons;
in our $dS_4$ example above, the longitudinal
graviton mode $s(x)$ for $m^2 < 0$ is an example of a tachyon which
is not a ghost. 

As we argued above, for scalars in flat space ghosts and tachyons
exhaust the list of physically distinct pathologies for weakly
coupled theories. This is less obvious for gravity. However if we
avoid regions of strong coupling and abstain from integrating
out degrees of freedom, there does not seem to be any reason to
impose additional positivity constraints. We will assume this for
our analysis.

There is one other possible source of ambiguity. Because gravity is a highly
nonlinear theory, one could imagine that there are physical setups
in which even the long--range solutions are intrinsically nonlinear.
In such a case ghosts of the linearized solutions are irrelevant,
since the linearized solution is not itself an approximation to
the real solution, even very far away from any source. It has been
argued that this may occur in the DGP model \cite{Gabadadze:2005qy}.
Note that the claim that nonlinearities of gravity are important
at cosmological scales is different from the claim that
matter nonlinearities are important at cosmological 
scales \cite{Gruzinov:2006nk};
the latter has also been suggested as a replacement for dark energy
in explaining late time accelerated 
expansion \cite{Kolb:2005da}. At any rate
we will not pursue this idea here, since we find the
linearized analysis of our models quite challenging enough!

\section{Warped Models with Induced Gravity\label{model}}

In this section we will review the main features of the two-brane
models introduced in \cite{Carena:2005gq,Bao:2005bv}, 
extending
the results to include the self--accelerating branch and the case
of positively curved $dS_4$ branes.
The model is described by the following action
\begin{eqnarray}
\label{eqn:ouraction}
S &=&
\int d^4x\,
\int^{L^-}_{0^+} dy
\sqrt{-G} \Big(4M^3 R - 2\Lambda\Big) \nonumber\\
&&+ \sum_i \int_{y=y_i} d^4 x
\sqrt{-g^{(i)}} (2M_i^2 \tilde{\cal R}^{(i)} - V_i)
+ 4M^3 \oint_{\partial \cal M} K \,.
\end{eqnarray}
This action represents a general warped
gravity setup with codimension one branes, written
in the interval picture. We have only displayed one
interval since physics on the second interval is
fixed by a $\mathbf{Z_2}$ symmetry. In (\ref{eqn:ouraction})
$M$ is the 5d Planck scale, $\Lambda = -24M^3k^2$
is the bulk cosmological constant giving a bulk curvature $k$,
the $M_i$ are the coefficients of brane-localized
curvatures $\tilde{\cal R}^{(i)}$,
the $V_i$ are brane tensions and $K$ is the
trace of extrinsic curvature.
As described in \cite{Carena:2005gq} we will use the
straight gauge formalism to keep the brane locations
fixed at $y=0^+$ and $y=L^-$, even in the presence of
linearized fluctuations of the metric. There are no
``brane-bending'' modes in a straight gauge.

The background solution can be written as
\begin{equation}
\label{eqn:metric}
ds^2=G_{MN}dx^Mdx^N=g_{\mu\nu} dx^\mu dx^\nu + dy^2,
\end{equation}
where $y$ is restricted to the interval $0<y<L$ and
\begin{equation}
g_{\mu\nu} = a^2(y) \gamma_{\mu\nu}\,,
\end{equation}
with $\gamma_{\mu\nu}$ the metric of $dS_4$ or $AdS_4$ with
constant curvature $12\mathcal{H}^2$.
The explicit form of the warp factor, $a(y)$, the 4d curvature constant
$\mathcal{H}^2$, and the coordinate size of the extra dimension $L$, 
depend on the input brane parameters that also
determine how the $AdS_5$ bulk is sliced. It is convenient to
define the brane parameters in terms of dimensionless quantities,
\begin{equation}
v_i=k M_i^2/M^3, \quad w_i=V_i/2kM^3 \, .
\end{equation}
The derived features of the model are determined by the following
combinations of input parameters:
\begin{equation}
T_i^\pm \equiv \frac{1}{v_i} \left(
-1 \pm \sqrt{1+v_i w_i/6+v_i^2} \right) \, .\label{Tpm}
\end{equation}
The choice of sign in (\ref{Tpm}) corresponds to the two
independent branches (per brane) of the background solutions to the
equations of motion.

\subsection{slicing the $\mathbf{AdS_5}$ bulk}

For a given set of five input parameters $k$, $v_0$, $w_0$, $v_L$,
$w_L$, the first question is to determine the possible $dS_4$ or $AdS_4$
slicings of the $AdS_5$ bulk. This is set by
whether the absolute value of $T_i$ is larger or smaller
than one. We show in Figure~\ref{regions:fig} the regions in the $(v,w)$
plane that correspond to $|T|>1$ with dark shade and $|T|<1$ in
light shade. In the unshaded regions $T$ is complex.
The left and right panels are for $T^+$ and $T^-$,
respectively. 
The curved boundaries correspond to the line 
$w=-6(1+v^2)/v$, the horizontal lines that separate the dark and light
areas correspond to $w=\pm 12$ (and $v$ smaller or larger than $\pm
1$, depending on the panel) and finally the slanted line on the left
panel separating the two light shades (with $T^+$ positive or
negative) corresponds to $w=-6 v$.

Simply put, in the dark shaded regions the branes are both
$dS_4$, in the light shaded regions they are both $AdS_4$. We
parametrize the warp factor differently in each case:

\begin{figure}[ht]  
\centerline{\includegraphics[width=0.45\textwidth]{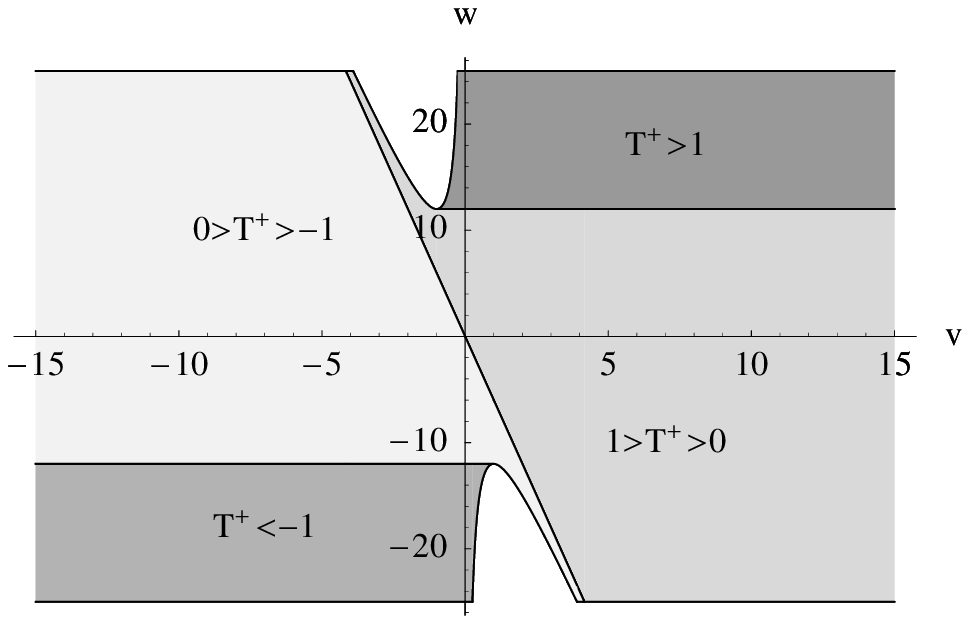}
\includegraphics[width=0.45\textwidth]{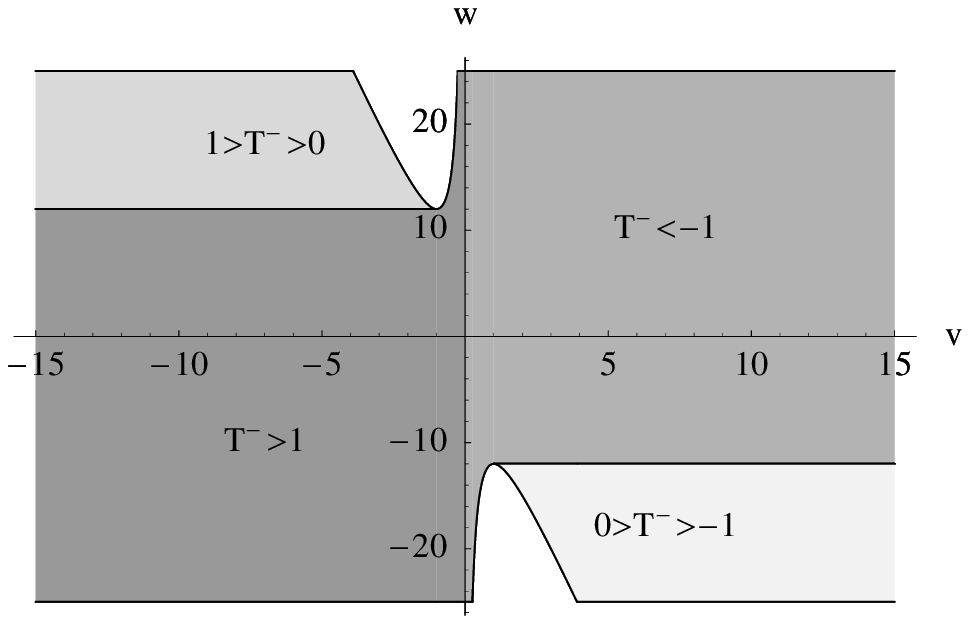}
}
\caption{
Regions of the $(v,w)$ plane corresponding to different background
solutions and signs of $T$. The left panel is for the 
$T^+$ branch and the right panel is for the $T^-$ branch. The two (lighter)
darker shades correspond to $(A)dS_4$ branes. In the unshaded regions
the branes are neither $AdS_4$ nor $dS_4$. 
}
\label{regions:fig}  
\end{figure}  
\begin{itemize}
\item $\mathbf{AdS_5/AdS_4}$: If $|T_i|<1$, then the branes are $AdS_4$.
The warp factor reads,
\begin{equation}
a(y)=\frac{\cosh k(y-y_0)}{\cosh k y_0},
\end{equation}
with the turning point $y_0$ and $L$ given by
\begin{equation}
T_0=\tanh k y_0, \quad T_L=\tanh k(L-y_0). \label{Ti:AdS}
\end{equation}
Note that the requirement $L>0$ implies
\begin{equation}
T_0+T_L>0. \label{Lpositive:AdS}
\end{equation}
Finally the brane curvature is given by ($\mathcal{H}^2
\equiv -H^2 <0$)
\begin{equation}
H=\frac{k}{\cosh k y_0}.
\end{equation}
\item $\mathbf{AdS_5/dS_4}$: If $|T_i|>1$, then the branes are $dS_4$.
The warp factor reads,
\begin{equation}
a(y)=-\frac{\sinh k(y-y_0)}{\sinh k y_0},
\end{equation}
with $y_0$ and $L$ given by
\begin{equation}
T_0=\coth k y_0, \quad T_L=\coth k(L-y_0).
\end{equation}
Note that if $0<y_0<L$, the warp factor vanishes at some point in the
bulk. This is a true singularity and therefore does not give
rise to an acceptable model. Thus we have to require either $y_0<0$
or $y_0>L$ for the case of $dS_4$ branes. This
requirement, together with $L>0$, implies the conditions 
\begin{equation}
T_0\cdot T_L <0, \quad T_0+T_L< 0.
\end{equation}
Finally the brane curvature is given by 
\begin{equation}
\mathcal{H}=\tilde{\epsilon}_0\frac{k}{\sinh k y_0},
\end{equation}
where $\tilde{\epsilon}_0$ is the sign of $T_0$. 
\end{itemize}

\subsection{standard branches and self--accelerating branches}
For fixed input brane parameters, we have two different
background solutions per brane, denoted by the sign choice $\epsilon_i=\pm$ 
in $T_i^{\epsilon_i}$ (not to be confused with $\tilde{\epsilon}_i$,
the sign of $T_i$ itself). These two branches exhibit
different physical features.
In the following, we will focus our attention to
the brane at $y=0$. The restriction $L>0$, in addition to
$y_0<0$ or $L<y_0$ for
$dS_4$ branes, will then imply restrictions on the parameters of the
brane located at $y=L$, such that not all combinations of
signs give sensible solutions. 

The choice of branch for
the brane at $y=0$, \textit{i.e.} the choice of $\epsilon_0$, 
corresponds to the standard and self--accelerating branches
for the cosmology of brane--induced gravity. 
In our
notation, the standard branch corresponds to $T_0^+$ whereas the
$T_0^-$ branch is the self--accelerating one. The reason is the
following. For fixed values of the brane parameters, $v_0, w_0$,
the brane curvature is larger for the self--accelerating branch than
for the standard one\footnote{The name self--accelerating is actually
a bit of abuse of language in these more general configurations as
there are regions of parameter space 
where the $T_0^-$ branch gives an $AdS_4$ background, which is never
accelerating. Nevertheless even in this
case there is a net positive contribution to the brane curvature as
compared with the standard branch.}. 
This is obvious for $|w_0|<12$, for which $T_0^+$
corresponds to $AdS_4$ solutions whereas $T_0^-$ gives a $dS_4$
brane (see Fig.~\ref{regions:fig}). 
In the rest of the parameter space, where we have either two $AdS_4$
or two $dS_4$ solutions, it is still true that
\begin{equation}
\mathcal{H}^2\big[T_0^+(v_0,w_0)\big] < 
\mathcal{H}^2\big[T_0^-(v_0,w_0)\big].
\end{equation} 
Furthermore, for positive $v_0$, the
Randall-Sundrum~\cite{Randall:1999ee} tuning $w_0=+12$
results in a flat brane (independently of the particular value of
$v_0$) for $T_0^+$ and an inflating brane for $T_0^-$, thus the names
standard and self--accelerating branches. This is the
exact warped analogue of the DGP tensionless brane.
In the following we will loosely use the term branch to denote the
choice of sign in either brane. However, as we said, it is the brane 
at $y=0$ that determines the standard or self--accelerating
nature of 
the solution, and therefore we will only apply these terms to
the brane at $y=0$.

\subsection{properties of the spectrum}
The properties of the physical spectrum were described for the
$AdS_4$ brane case in \cite{Carena:2005gq,Bao:2005bv}, and can
be easily extended to the $dS_4$ case. The spectrum is obtained by
computing the action at the quadratic level for small perturbations
around the background metric,
\begin{equation}
G_{MN}\to G_{MN}+h_{MN}.
\end{equation}
This can be done in a straight gauge, for which the branes are
straight and at fixed positions, $y=0^+,L^-$, and $h_{\mu 4}(x,y)=0$. A
further gauge--fixing exhibits the radion, 
\begin{equation}
h_{44}(x,y)=F(y) \psi(x),
\end{equation}
where $F(y)$ is a fixed but arbitrary function (part of the gauge choice)
and $\psi(x)$ is a
four-dimensional scalar. The components of $h_{\mu\nu}(x,y)$ can
then be shown to have the following 4d physical degrees of freedom:
\begin{itemize}
\item \textbf{massless modes}: there is a graviton
zero mode, $B_{\mu\nu}(x)$ (2 degrees of freedom) 
and the above mentioned radion, $\psi(x)$. 
\begin{equation}
h_{\mu\nu}= a^2(y) B_{\mu\nu}(x)+ 
a^2 {\mathcal Y}_1(y) \tilde{\nabla}_\mu \tilde{\nabla}_\nu \psi 
+ g_{\mu\nu} {\mathcal Y}_2(y) \psi \,. 
\end{equation}
where the $y$ dependence in the radion piece is given by
\begin{equation}
\mathcal{Y}_1(y) = \chi\frac{k}{\mathcal{H}^2}(z-1)^2 - {\mathfrak
F}\,,\quad 
\mathcal{Y}_2(y) =\chi k (z-1)^2 + \frac{a'}{a} {\mathcal F}\,.
\label{eqn:y1y2}
\end{equation}
In this last equation we have defined the variable $z$
\begin{equation}
z= \left \{ 
\begin{array}{ll} 
\tanh k(y-y_0), \quad &\mathcal{H}^2<0, \\
\coth k(y-y_0), \quad &\mathcal{H}^2>0. 
\end{array} \right.
\end{equation}
 Note that we have $T_i=\theta_i z(y_i)$, with
$-\theta_0=\theta_L=1$. $\chi$, $\mathcal{F}$ and $\mathfrak{F}$ are
remnants of the residual gauge freedom (related to $F(y)$) 
whose detailed form is not important for the discussion below.
\item \textbf{massive modes}: there is a tower of massive gravitons 
(5 degrees of freedom each)
\begin{equation}\label{eqn:massiveh}
h_{\mu\nu} = \sum_{q} b^{(q)}_{\mu\nu}
= \sum_{q}{\mathcal Y}^{(q)}(y) B^{(q)}_{\mu\nu}(x)\,,
\end{equation}
where $q$ labels the mass, $q =m^2/\mathcal{H}^2$, and
\begin{eqnarray}\label{eqn:massiveysol}
\mathcal {Y}^{(q)}(y) &=& P_{(-1+\sqrt{9- 4q})/2}^{-2}(z) 
- \frac{a^{(q)}_0}{b^{(q)}_0} Q_{(-1+\sqrt{9- 4q})/2}^2(z) \nonumber\\
&=& P_{(-1+\sqrt{9- 4q})/2}^{-2}(z) 
- \frac{a^{(q)}_L}{b^{(q)}_L} Q_{(-1+\sqrt{9- 4q})/2}^2(z) \,,
\end{eqnarray}
where the $P$'s and $Q$'s are associated 
Legendre functions.
The mass spectrum of modes is determined by solving the determinant equation, 
\begin{equation}\label{eqn:deteq}
a_0 b_L - a_L b_0  = 0\; ,
\end{equation}
with
\begin{eqnarray}
a_i &=& 2\theta_i
(1-T_i^2)\frac{d}{dz}P_{-\frac{1}{2}+\frac{1}{2}\sqrt{9-4q}}^{-2}(z)  
\Big|_{z=\theta_i T_i} 
- \{v_i q (T_i^2-1) + 4T_i\} 
P_{-\frac{1}{2}+\frac{1}{2}\sqrt{9-4q}}^{-2}(\theta_i T_i) \,,\\
b_i &=& 2\theta_i
(1-T_i^2)\frac{d}{dz}Q_{-\frac{1}{2}+\frac{1}{2}\sqrt{9-4q}}^2(z)  
\Big|_{z=\theta_i T_i} 
- \{v_i q (T_i^2-1) + 4T_i\} 
Q_{-\frac{1}{2}+\frac{1}{2}\sqrt{9-4q}}^2(\theta_i T_i) \,.
\end{eqnarray}
\end{itemize}
All these equations are valid for both signs of the brane curvature,
provided the right definition of $z$, the warp factor, $a(y)$, and the
sign of the brane curvature $\mathcal{H}^2$ are used.\footnote{
The price for this convenient notation is that $q$ as defined here
is $-q$ as defined in \cite{Carena:2005gq}.}
Performing the integration over the extra dimension in the
quadratic terms for the different modes we obtain the corresponding
kinetic coefficients (normalization constants). For the
massless modes:

\begin{eqnarray}
{\mathcal C}_g^{(0)}&=&
\frac{T_0^2-1}{k}\sum_i\bigg[-k y_i 
+ \frac{T_i+v_i}{T_i^2-1} \bigg]\,, 
\label{Cg0} \\
{\mathcal C}_\psi &=& -\frac{3\chi^2{\mathcal H}^2}{2k} \sum_i 
\bigg[T_i+\frac{ v_i + T_i}{1+v_i T_i}
\bigg]\,,\label{Cpsi}
\end{eqnarray}
and for
the massive modes: 
\begin{equation}
{\mathcal C}^{(q)}_g = -\frac{k}{{\mathcal H}^2} \Big( 2\int_{-T_0}^{T_L} 
{\mathcal Y}^{(q)}{}^2 \mathrm{d} z 
+ \sum_i [v_i (1-z^2) {\mathcal Y}^{(q)}{}^2]_{z=\theta T_i} \Big)\,.
 \label{Cgq}
\end{equation}
If any of these coefficients becomes infinite, the corresponding mode
is not normalizable and decouples from the spectrum. If it vanishes,
then we are in a region of strong coupling.
If it becomes negative, then the mode is a ghost. 

\subsection{longitudinal graviton modes in $\mathbf{dS_4}$}

As already mentioned, for $dS_4$ the kinetic terms of
some massive graviton modes have an intrinsic mass-dependent
kinetic coefficient already at the 4d level, in addition to
the overall coefficient ${\mathcal C}^{(q)}_g$ that they inherit
from the Kaluza-Klein (KK) decomposition. Thus in this case
we need to compute this intrinsic coefficient before deciding
if the mode is a ghost.

To compute these coefficients, we need an explicit orthogonal
decomposition of the five ``helicities'' of each massive
4d graviton $B_{\mu\nu}^{(q)}(x)$. 
The
appropriate orthogonal decomposition turns out to be:
\be\label{eqn:ourodec}
B_{00}^{(q)}(x) &=& f^{-3/2}\nabla_k^2 s^{(q)}(x) \; ,\nn\\
B_{0i}^{(q)}(x) &=& 
f^{1/2}\nabla_k^2 v_i^{(q)}(x) + f^{-2}\partial_0\partial_i 
(f^{1/2}s^{(q)}(x))
\; ,\\
B_{ij}^{(q)}(x) &=& f^{1/2}b_{ij}^{(q)}(x) 
+ f^{-1}\partial_0(f^{3/2}(\partial_i v_j^{(q)}(x) + \partial_j v_i^{(q)}(x)))
\nn\\
&&\hspace*{20pt}
+\frac{1}{2}f^{1/2}P_{ij}\nabla_k^2 s^{(q)}(x)
+(\delta_{ij} - \frac{3}{2}P_{ij})\partial_0(\partial_0+H) 
(f^{1/2}s^{(q)}(x))
\; ,\nn
\ee 
where 
\be
P_{ij} \equiv \delta_{ij} - \frac{\partial_i\partial_j}{\partial_k^2}
\ee
is a transverse projection operator in the flat 3d sense,
and $\nabla_i^2 = \partial_i^2/f(t)^2$ is the 3d Laplacian
defined already in the previous section.
In (\ref{eqn:ourodec}) the $b_{ij}^{(q)}(x)$ are traceless-transverse
in the flat 3d sense (2 degrees of freedom), corresponding to
the helicity $\pm 2$ modes of the massive gravitons in the KK
tower. The $v_i^{(q)}(x)$ are transverse vectors in the flat
3d sense (2 degrees of freedom), corresponding to
the helicity $\pm 1$ modes of the massive gravitons in the KK
tower. The $s^{(q)}(x)$ are the longitudinal modes of the
massive gravitons.

Substituting the ansatz (\ref{eqn:ourodec}), it is straightforward
but tedious to show that the full 5d bulk equations of motion reduce
to a set of constraint equations which are automatically satisfied,
together with the following dynamical 4d equations of motion in
the de Sitter background:
\be
0 &=& \left( 
-\partial_0^2 + \nabla_i^2 - ( m^2 -\frac{9}{4}\mathcal{H}^2)
\right)b_{ij}^{(q)}(x) \; ,\nn\\
0 &=& \nabla_j^2\left( 
-\partial_0^2 + \nabla_i^2 - ( m^2 -\frac{9}{4}\mathcal{H}^2)
\right)v_{i}^{(q)}(x) \; ,\\
0 &=& \nabla_j^2\left( 
-\partial_0^2 + \nabla_i^2 - ( m^2 -\frac{9}{4}\mathcal{H}^2)
\right)s^{(q)}(x) \; .\nn
\ee
Solutions to the equation of motion for the vectors $v_{i}^{(q)}(x)$
and scalars $s^{(q)}(x)$
appear to be defined only up to an arbitrary harmonic function
annihilated by $\nabla_i^2$, but a residual 4d general coordinate
invariance removes this ambiguity.

We can also substitute our orthogonal decomposition (\ref{eqn:ourodec}) into
the effective 4d quadratic action, to read off the intrinsic
mass-dependent kinetic coefficients. Since the decomposition
involves derivatives, this is really only meaningful with the
derivatives evaluated on-shell. The interesting case is for the
longitudinal modes, where after another tedious calculation
we obtain the effective action shown in (\ref{eqn:lmodekin})
for each longitudinal graviton mode in the KK tower.
Thus, even for values of $q= m^2/\mathcal{H}^2$ 
such that ${\mathcal C}^{(q)}_g > 0$, the longitudinal graviton
mode will be a ghost if $0 < q < 2$.

\section{Ghosts in Models with Induced Gravity\label{spectrum}}

In this section we will discuss the spectrum of our models in the
search of regions that are free of ghost and tachyonic instabilities.
This is done by studying the solutions of the eigenvalue equation 
for the massive modes (\ref{eqn:deteq}) as well as the kinetic
coefficients of these and the graviton zero mode and radion,
(\ref{Cg0}-\ref{Cgq}). 
It is useful to 
separate the discussion for $AdS_4$ and $dS_4$ branes, because of the
different behavior of the longitudinal component of massive gravitons
in the latter.
Thus, in an $AdS_4$ space, we only need to look at
the kinetic coefficients of the different modes, ghosts being uniquely
determined by the negative sign of their kinetic terms, whereas in
$dS_4$ space, the longitudinal component of a massive graviton with
positive overall kinetic coefficient but with mass
$0<m^2<2H^2$ is a ghost, becoming a (non-ghost) tachyon for
$m^2<0$.

\subsection{negative curvature: $AdS_4$ branes}

The presence of ghosts for $AdS_4$ branes was discussed, for
the standard branch, in \cite{Bao:2005bv}. In this
section we will review the results and generalize them 
to the self--accelerating
branch. 
Let us start discussing the simpler case of the
graviton zero mode.
Recalling the definition of
$T_i$ in the $AdS_4$ background (\ref{Ti:AdS}), we can write the
kinetic coefficient for the graviton zero mode as
\begin{equation}
\mathcal{C}_g^{(0)AdS_4}=\frac{1-T_0^2}{k} \sum_i 
\left[\tanh^{-1}T_i + 
\frac{T_i+v_i}{1-T_i^2} \right].
\end{equation}
For $AdS_4$ the global term outside the sum is positive so we only
need to look at the sign of the sum. Let us define 
each term in the sum as
\begin{equation}
\bar{\mathcal{C}}^{(0),i}_{g} \equiv \tanh^{-1}T_i+\frac{v_i+T_i}{1-T_i^2}.
\end{equation}
$\bar{\mathcal{C}}^{(0),i}_g$ is a growing function of
$w_i$, for fixed $v_i$. Let us now consider its value at the
boundaries of the $AdS_4$ region in the $(v,w)$ plane (light shaded
areas in Fig.~\ref{regions:fig}). 
In the limit $v_i \to \pm \infty$ along the curve 
$w_i=-6(1+v_i^2)/v_i$ we get $\mathcal{C}^{(0),i}_g=\pm\infty$. 
As we move along that curve towards the points
$v_i \to \pm 1$ (and $w_i=\mp 12$, respectively),
it goes to $\mp \infty$. This is independent of which branch we have
chosen for the branes. The rest of the boundaries, however, depend on
whether we have $T^+_i$ or $T^-_i$ (see Fig.~\ref{regions:fig}).
For $T^+_i$, $\bar{\mathcal{C}}^{(0),i}$ stays $+\infty$ for $w_i=+ 12$ and $v_i>-1$ and
$-\infty$ for $w_i=-12$ and $v_i<1$ whereas for $T^-_i$
it is
$-\infty$ for $w_i=+12$, $v_i<-1$ and $+\infty$ for $w_i=-12$ and
$v_i>1$. In summary, for any branch
it goes all the way to $-\infty$ for negative $v_i$ and all the way to
$+\infty$ for positive $v_i$. Furthermore it can be made as negative
(positive) as one wants for sufficiently large negative (positive) 
$v_i$. Thus, it is clear that 
independent of the value of $\bar{\mathcal{C}}^{(0),L}_g$ (provided it
remains finite)
there will always be a curve
splitting the 
$(v_0,w_0)$ plane in two parts,
such that to the left of the curve 
the graviton zero mode is a ghost and to the right of the curve
it is not.
The case of the radion
is more complicated so rather than giving analytic expressions
describing the different behaviors, we prefer to show, directly in the
figures, the relevant cases for the different choices of branches.

There is also an extra source of instability in these models,
namely the possible existence of tachyons in the spectrum. 
A thorough study of tachyons 
for the case of $AdS_4$ branes in
the standard branch was done in \cite{Bao:2005bv}, 
and the techniques to study them 
apply mostly unchanged to the
self--accelerating branch.
Recall that in our
notation, due to $\mathcal{H}^2(AdS_4)<0$, tachyonic solutions are
characterized by $q>0$.
It proves useful to study the behavior of the quantity
\begin{equation}
\mathfrak{D}\equiv \frac{a_0 b_L-b_0 a_L}{q(1-T_0^2)(1-T_L^2)},
\label{mathfrak:D}
\end{equation}
that has the same massive solutions as our original eigenvalue
equation, but we have explicitly removed the zero mode solution,
$q=0$. 
The behavior of this function at $q=0$ and $q \to \infty$
is, respectively \cite{Bao:2005bv},  
\begin{equation}
\mathfrak{D}(q=0)=2 \frac{k}{T_0^2-1} \mathcal{C}_g^{(0)},
\end{equation}
and
\begin{equation}
\mathrm{sign}\big[ \mathfrak{D}(q\to \infty) ] = 
\mathrm{sign}\big[v_0 v_L (\cos^{-1}T_0 + \cos^{-1} T_L - \pi)]. 
\end{equation}
Thus, if the signs of both are different, there exists an odd number
(therefore at least one) of solutions with positive $q$, \textit{i.e.}
tachyons. If the signs of both are the same, then there is either no
tachyon or an even number of them (that last distinction can be
resolved numerically). The condition $L>0$ gives, for $AdS_4$
branes, $T_0+T_L>0$ which in turn implies
\begin{equation}
\cos^{-1} T_0 + \cos^{-1}T_L - \pi <0. 
\end{equation}
It is then clear that in order to have an even number of tachyons
(usually zero), we need
\begin{equation}
\mathrm{sign}~v_0 v_L = \mathrm{sign}~ \mathcal{C}_g^{(0)}>0,
\end{equation}
where the last inequality is for the phenomenologically relevant region
in which the graviton zero mode is not a ghost. 

\begin{figure}[ht]
\includegraphics[width=.45\textwidth]{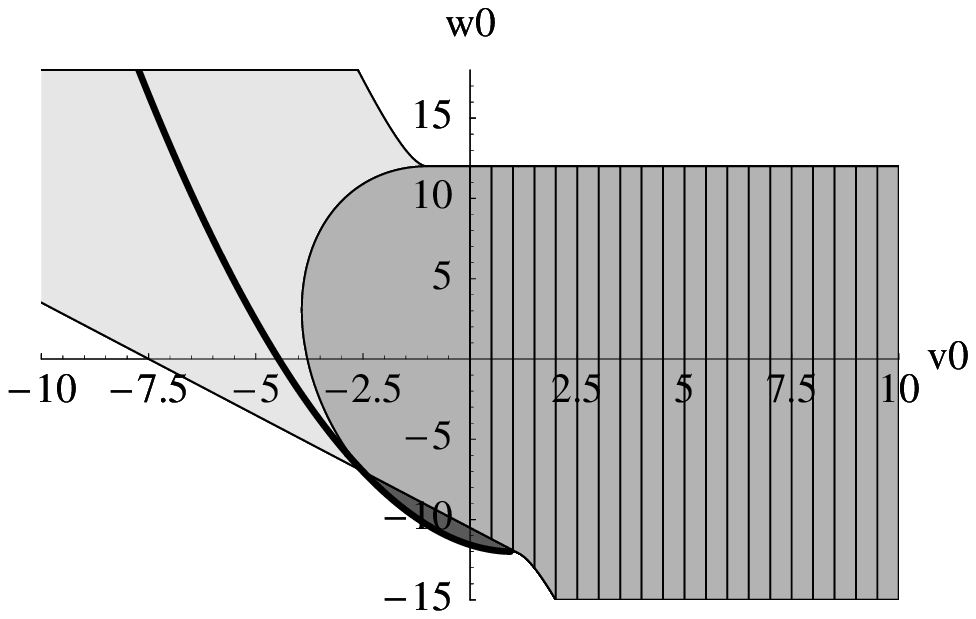}
\includegraphics[width=.45\textwidth]{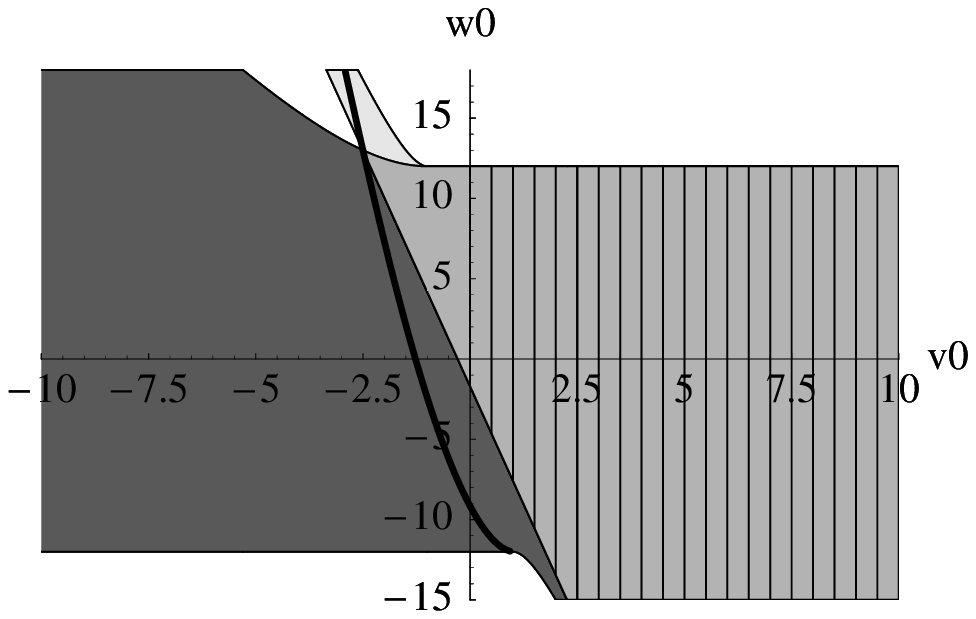}
\caption{\label{TpTp:AdS} Ghost and tachyon--free 
region for the $(T_0^+,T_L^+)$
branch. The different shades
correspond to the following
regions: $L>0$ (light), $\mathcal{C}_\psi>0$ (dark) and the intersection of
both (intermediate). The hatched area corresponds to the region where
$\mathcal{C}_\psi>0$, $L>0$ and there are not tachyons in the spectrum.
The region to the right of the thick solid line
corresponds to $\mathcal{C}_g^{(0)}>0$. Thus, the hatched area to the right of
that line is the ghost and tachyon--free region for this
configuration. 
In the left panel $v_L=2.5$, $w_L=7$ whereas in the right panel we use  
$v_L=2.5$ and $w_L=-13$.} 
\end{figure}
In Fig.~\ref{TpTp:AdS} we show the case of the $(T_0^+,T_L^+)$ branch for
two possible behaviors, depending on which choice of parameters we
make for the brane at $y=L$.  We represent three different shades in
the figure, the light 
one is the region of parameter space for which $L>0$, the dark one is
the region in which the radion is not a ghost and the intermediate
shade is the region for which both conditions are satisfied. 
That latter area is hatched if no tachyons are present in the
spectrum. 
In these examples we have $v_L=2.5>0$ and therefore the condition for
no tachyons in the spectrum demands $v_0>0$. 
Finally, the thick solid line represents the points for
which $\mathcal{C}_g^{(0)}=0$. 
Therefore the hatched region to the right of the solid line is
ghost and tachyon free. (It can be checked numerically that the massive
gravitons have positive kinetic coefficients in that region.)
Note that there are also regions of parameter space with
positive and small $v_0$ that are not allowed.

\begin{figure}[ht]
\includegraphics[width=.45\textwidth]{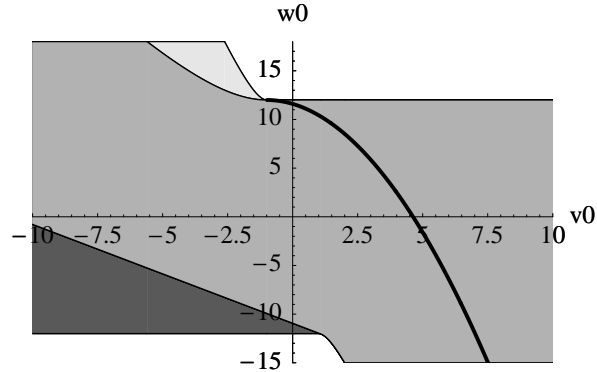}
\caption{\label{TpTm:AdS} Ghost-free region for the $(T_0^+,T_L^-)$ 
branch, with $w_L>12$. We have taken $v_L=-4$ and $w_L=15$. 
The region to the right of the thick
solid line corresponds to $\mathcal{C}_g^{(0)}>0$. 
The shades correspond to the following
regions: $L>0$ (light), $\mathcal{C}_\psi>0$ (dark) and the intersection of
both (intermediate). The ghost--free region has tachyons in the
spectrum.}
\end{figure}

In Fig.~\ref{TpTm:AdS}, we consider the case of the $(T_0^+,T_L^-)$
branch.  
In this case there is a large
region where the radion is not a ghost for both values of the sign of
$v_0$, provided $w_L>12$ (considering values of
$w_L<-12$ yields a very small region for which  the radion is not a ghost).
In the figure, we have chosen $v_L=-4$ and $w_L=15$. 
The shades have the same meaning as in the previous figure. 
From Fig.~\ref{regions:fig}, it follows that $w_L>12$ implies $v_L<0$
for $T_L^-$, thus the region to the
right of the graviton zero mode line with $v_0>0$ 
has an odd number of tachyons. 
A numerical analysis shows that the small region to the right of
the graviton zero mode line but with $v_0<0$ has two
tachyons. Thus, this solution does not have any region that is ghost
and tachyon free (no hatched area).

\begin{figure}[ht]
\includegraphics[width=.45\textwidth]{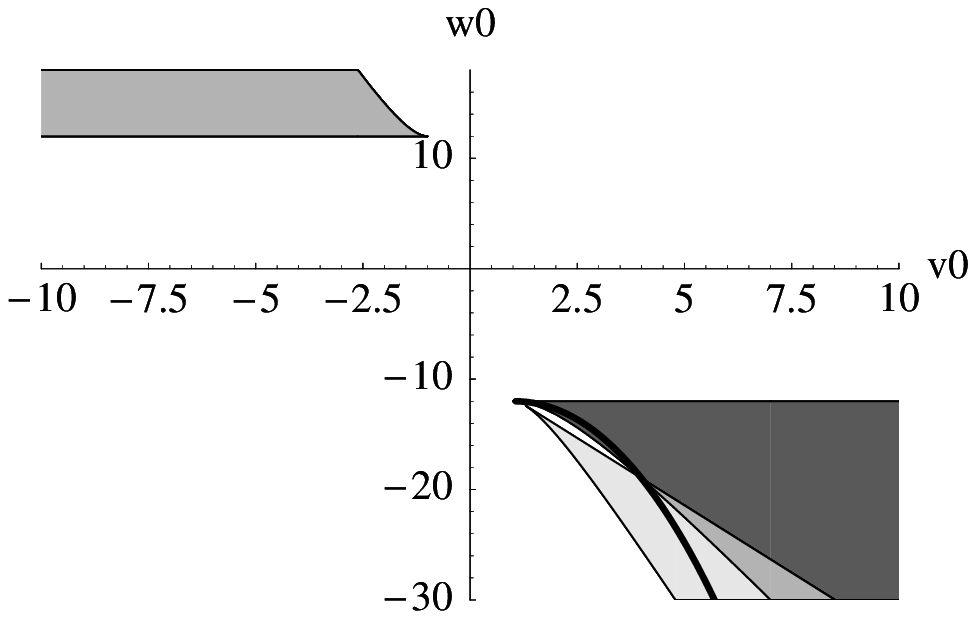}
\includegraphics[width=.45\textwidth]{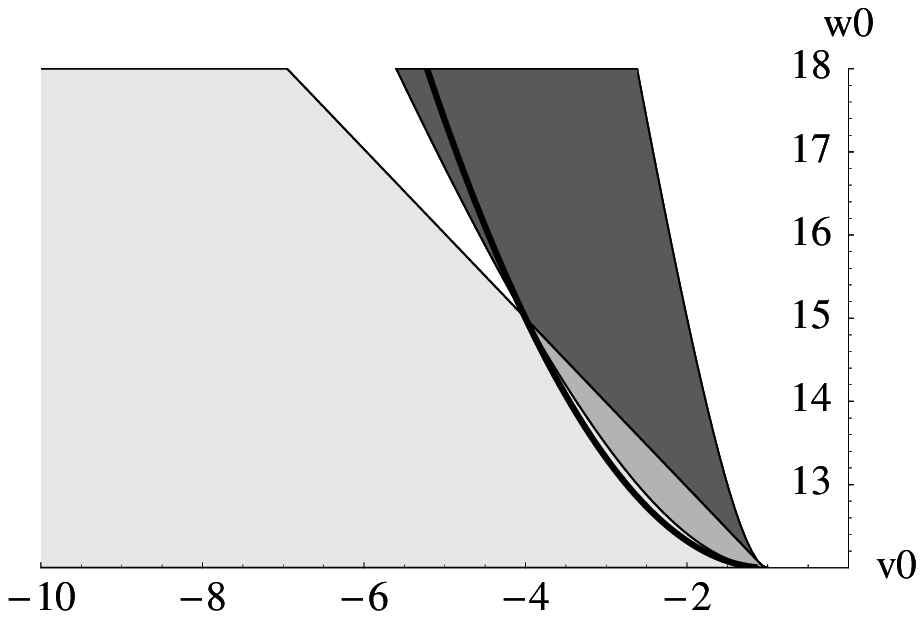}
\caption{\label{TmTm:AdS} Ghost-free region for both
the $(T_0^-,T_L^-)$ branch. The shades 
correspond to the following
regions: $L>0$ (light), $\mathcal{C}_\psi>0$ (dark) and the intersection of
both (intermediate). The region to the right of the thick black line
corresponds to $\mathcal{C}_g^{(0)}>0$.
In the left panel we have $v_L=-4$, $w_L=19$ whereas for the right panel
$v_L=4$ and $w_L=-15$.} 
\end{figure}

Let us consider now the cases of self--accelerating branches:
$(T_0^-,T_L^-)$ and $(T_0^-,T_L^+)$.
In Fig.~\ref{TmTm:AdS} we show the case
$(T_0^-,T_L^-)$. 
There is a large ghost--free region in the case that $w_L>12$ whereas
it shrinks to almost nothing if $w_L<-12$.
From Fig.~\ref{regions:fig}, it follows that $w_L>12$ implies $v_L<0$
(left panel in the Fig.~\ref{TmTm:AdS}) and $w_L<-12$ implies $v_L>0$
(right panel).
The ghost-free areas have,
respectively, $v_0>0$ and $v_0<0$ and therefore at least one tachyon. 
Thus in this branch there are no ghost and tachyon-free regions.
In the $(T_0^-,T_L^+)$ branch, 
there is a large ghost--free region with
$w_0>12$ (and therefore $v_0<-1$), 
quite independent of the values of $v_L$ and $w_L$. This ghost--free
region, however, always has tachyons and therefore there are no 
ghost and tachyon--free regions in this case, either. We do not
present a figure for this case since
it is quite similar to the left upper
region of the left panel in Fig.~\ref{TmTm:AdS}.

In summary, in the $AdS_4$ case, only the $(T_0^+,T_L^+)$ branch with
$v_0>0$ and $v_L>0$ has ghost and tachyon--free solutions.

\subsection{positive curvature: $dS_4$ branes}

We now consider the phenomenologically more relevant case of inflating
branes. This corresponds to a choice of brane parameters such that
$|T_i|>1$. Recall that the absence of a true singularity in the bulk
forces us to choose opposite signs for $T_0$ and $T_L$. This leaves
eight different regions of parameter space: four choices of
combinations $T_0^\pm$, $T_L^\pm$ times two choices of the sign of
$T_0$. 
We have emphasized above the possible existence of a
new source of ghosts for inflating branes, the longitudinal component
of a massive graviton with mass $0< m^2 < 2 H^2$. 
There is a deep connection, already present in the DGP model, 
between the radion and the massive graviton
ghosts, that can be characterized by the following identity,
\begin{equation}
{\mathfrak D}(q=2)
=\mathrm{sign}(T_0) \,
\frac{8}{3}(1+v_0 T_0)(1+v_L T_L) \bar{\mathcal
C}_\psi\,,  \label{eqn:dq2=cpfntl} 
\end{equation} 
where we defined $\mathfrak{D}$ in (\ref{mathfrak:D}) and
$\bar{\mathcal{C}}_\psi \equiv k \mathcal{C}_\psi/\chi^2\mathcal{H}^2$ has the same sign
as $\mathcal{C}_\psi$ for $dS_4$. 
This identity states that at the boundary between the regions where the
radion is or is not a kinetic ghost, there is always a massive graviton
with $m^2=2H^2$, thus also at the boundary of having a ghost
longitudinal component. 
In the DGP model, the dependence on the
brane parameters is such that, as we cross from the region where the
radion is a ghost to the region where it is not, a massive graviton
goes from mass squared larger than $2 H^2$ to mass squared smaller
than that, so that there is always a ghost in the spectrum. In our
case, the richer parameter space allows for different behaviors.   
Let us denote by $s\equiv(v_i^\ast,T_i^\ast)$ a point for which
$\bar{\mathcal{C}}_\psi(s)=\mathfrak{D}(q=2)=0$. 
The variation of the solution of the equation 
\begin{equation}
\mathfrak{D}(q)=0,
\end{equation}
around $s$ for fixed $v$ leads to 
(see Appendix~\ref{detailed:calculations} for details) 
\begin{equation}
\left. \frac{\delta q}{\delta T_0} \right|_s
= \frac{8}{3} \frac{1}{B(v_0,w_0,v_L,w_L)} 
\left.\frac{\partial \bar{\mathcal{C}}_\psi}{\partial T_0}
\right|_{T_0^\ast},
\end{equation}
where $B(v_0,w_0,v_L,w_L)$ is a function that is \textit{negative} in
the region of parameter space for which $L>0$ and \textit{positive}
otherwise. 
Thus, if $s$ is in the physical region with $L>0$, 
the slopes of the solution of the massive
graviton and the kinetic coefficient of the radion have opposite signs,
which means that \textit{near the boundary defined by
$\bar{\mathcal{C}}_\psi=0$,  
either the radion or a massive graviton is always a ghost}. If, on the
other hand the point $s$ lies outside the physical region ($L<0$)
there is potentially a dramatic effect on the physical spectrum. In
this case,
contrary to what happens in the DGP model, there will be regions where
neither the radion, nor a massive graviton is a ghost.
We can study the presence of massive gravitons with $q<2$ in a way
similar to the study of tachyons in the $AdS_4$ case.
The behavior of $\mathfrak{D}(q)$ in the limit $q\to -\infty$ 
reads (see Appendix~\ref{detailed:calculations})
\begin{equation}
\mathrm{sign}[\mathfrak{D}(q\to -\infty)] = \mathrm{sign}[T_0 v_0
v_L].
\end{equation}
Thus, the sign of the determinant equation at minus
infinity changes when either $v_0$ or $v_L$ goes through zero.
Using (\ref{eqn:dq2=cpfntl}) and recalling the definition of $T_i$
we also have
\begin{equation}
\mathrm{sign}[\mathfrak{D}(q=2)]=\mathrm{sign}[T_0] \epsilon_0 \epsilon_L,
\quad (\mbox{for }\bar{\mathcal{C}}_\psi>0),
\end{equation} 
where $\epsilon_{0,L}$ are the sign choices (branches) for
$T_{0,L}^{\epsilon_{0,L}}$. A comparison of the sign of the
determinant at both points shows that, excluding the graviton zero
mode, there is an even (possibly zero)
number of modes with $q<2$ (\textit{i.e.} either ghosts or tachyons) 
if $\mathrm{sign}[v_0v_L]=\epsilon_0 \epsilon_L$ and an odd number of
them (thus at least one) otherwise.
In order to classify the different behaviors, it helps recalling
that $T_i^-$ has the opposite sign to $v_i$ (see
Fig.~\ref{regions:fig}, right panel), 
whereas $T_i^+$ has the same sign as $v_i$ everywhere except
in a small wedge, where it has the opposite sign
(Fig.~\ref{regions:fig}, left panel).

Let us start our discussion with the cases for which the line
$\bar{\mathcal{C}}_\psi=0$ is always outside the physical region.
As explained in Appendix~\ref{detailed:calculations}, this occurs in 
the $(T_0^-,T_L^+)$ branch for positive $T_0$ and the $(T_0^+,T_L^-)$
branch with $T_0$ negative or with $T_0$ positive and 
$(T_L-1)(2-v_L+v_L T_L)/(1+v_L T_L)>-4$. In
the latter case the radion is always a ghost whereas in the former two
it is never a ghost in the physical region of parameter space. These are
potentially very interesting cases, since, contrary to what happens
in the DGP model, they include a
self--accelerating solution with regions where neither the radion nor
the longitudinal component of a massive graviton is a ghost.
In this case we have
$\epsilon_0 \epsilon_L=-1$ and therefore the regions that satisfy
$\bar{C}_\psi>0$ with $v_0 v_L>0$
have an odd number of modes with $q<2$. 
This leaves the wedge of the brane in the $(+)$ branch as
the only possible ghost-free region.
The massive graviton/radion system is
actually ghost free in these regions. Unfortunately, having one of the
two $v_i$ negative makes the graviton zero mode a ghost.
\begin{figure}[ht]
\includegraphics[width=.45\textwidth]{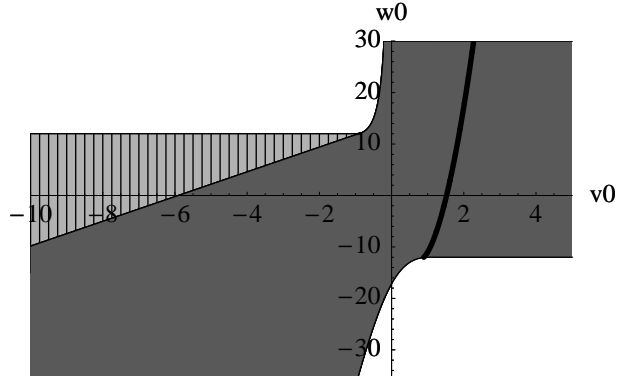}
\caption{\label{TmTp:dS} Ghost and tachyon-free region for the 
$(T_0^-,T_L^+)$ branch. We have chosen $v_L=0.05$ and $w_L=-13$. 
The shades 
correspond to the following
regions: $\mathcal{C}_\psi>0$ (dark) and the intersection of
$\mathcal{C}_\psi>0$ with $L>0$ and $y_0$ such that there are no
singularities in the background (light). 
The region to the right of the thick black line
corresponds to $\mathcal{C}_g^{(0)}>0$.
The hatched region corresponds to the physical
region where the radion is not a ghost and there are no modes with $q<2$.
In this case the only hatched region is to the left of the thick solid
line and therefore there is no ghost and tachyon free region for this
branch.}
\end{figure}
An example of this case is shown in Fig.~\ref{TmTp:dS}, where we have
chosen the $(T_0^-,T_L^+)$ branch with $v_L=0.05$ and $w_L=-13$ (thus
the brane in the $(+)$ branch living in the small wedge). 
The shaded region of the
figure represents the area where the radion is not a ghost, outside
the physical region (dark shade) and inside the physcal region (light
shade), \textit{i.e.} $\mathcal{C}_\psi>0$, $L>0$ and $y_0$ outside the
interval $[0,L]$. The curved lower boundary of the dark region represents the
line $\mathcal{C}_\psi=0$, which is outside the physical region and
therefore, above it, neither the radion nor the massive graviton is a
ghost. The hatched area is the physical region in which the radion is
not a ghost and all massive gravitons have $q>2$. This region is
to the left of the thick solid line and therefore the graviton zero
mode is a ghost.

\begin{figure}[ht]
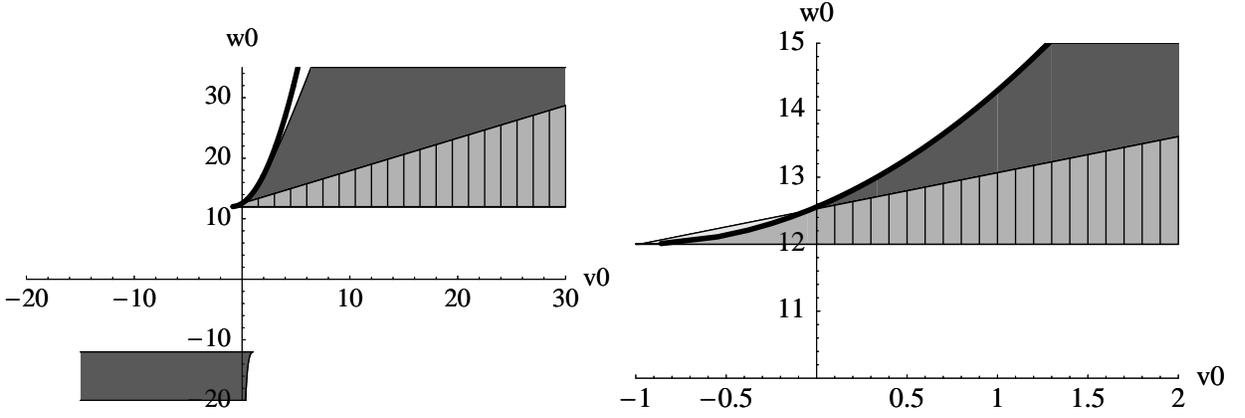

\includegraphics[width=.45\textwidth]{TpTpvLeq0p05wLeqneg12p5.epsi}
\includegraphics[width=.45\textwidth]{TpTpvLeq0p05wLeqneg12p5detail.epsi}
\caption{\label{TpTp:dS} Ghost and tachyon-free region for the 
$(T_0^+,T_L^+)$ branch. We have chosen $v_L=0.05$ and $w_L=-12.5$. The
right panel is just a zoomed in view of the same plot.
The shades 
correspond to the following
regions: $L>0$ and $y_0$ such that there are no singularities in the
background (light), 
$\mathcal{C}_\psi>0$ (dark) and the intersection of
both (intermediate). 
The region to the right of the thick black line
corresponds to $\mathcal{C}_g^{(0)}>0$.
The hatched region corresponds to the physical
region where the radion is not a ghost and there are no modes with $q<2$.
Thus the hatched region to the right of the thick solid line is the
ghost and tachyon free region.} 
\end{figure}
Let us now consider the case for which 
the line $\bar{\mathcal{C}}_\psi=0$ can be
inside the physical region so that, \textit{in the neighborhood} 
of such line, 
either the radion or the longitudinal
component of a massive graviton is a ghost. 
This happens 
for the $(T_0^-, T_L^-)$, $(T_0^+, T_L^+)$ branches and for
the $(T_0^-, T_L^+)$ branch for negative $T_0$ or the $(T_0^+, T_L^-)$
branch for positive $T_0$, provided 
$(T_L-1)(2-v_L+v_L T_L)/(1+v_L T_L)<-4$. 
The condition $T_0\cdot T_L <0$ immediately tells us that there is an odd
number of modes with $q<2$ for the $(T_0^-,T_L^-)$ branch, since
$v_0\cdot v_L <0$ while $\epsilon_0 \epsilon_L=+1$. 
Similarly, the
$(T_0^+,T_L^+)$ branch has an odd number of modes with $q<2$ 
unless one (and only one) of the
two branes lives in such small wedges. Numerical analysis then shows
that the region with both $v_0,v_L>0$ is ghost and tachyon--free 
whereas there are
two modes with $q<2$ if they are both negative. An example of this is
shown in Fig.~\ref{TpTp:dS}, where we have chosen $v_L=0.05$ and
$w_L=-12.5$ (thus the $L$ brane in the small wedge). The color coding
is the same as in the previous figure. This time, there is a 
hatched area to the right of the thick solid line (with $v_0>0$)
that is ghost and tachyon--free. In the right panel we show a closer view of
the region around $v_0=0$ that shows an area with intermediate shade 
to the right
of the thick solid line for which neither the radion nor the graviton
zero mode are ghosts but there is one mode with $q<2$ (intermediate
shade unhatched
region with $v_0<0$).
For the $(T_0^+,T_L^-)$ and $(T_0^-,T_L^+)$ branches we have
$\epsilon_0 \epsilon_L=-1$, thus again only the
wedge of the brane in the $(+)$ branch might give ghost free
solutions. However,
numerical analysis always shows two modes with $q<2$ in the region
where the radion is not a ghost.

Our discussion so far has assumed a finite $L$. The different
properties we have described naturally
generalize to an infinite extra dimension 
by taking the limit $L \to \infty$. 
Note however that, 
in the $dS_4$ case, if $y_0>0$, there is a different \textit{infinte
extra dimension} limit we can take, namely $L \to y_0$. This
corresponds to an infinite extra dimension in conformal coordinates,
for which
\begin{equation}
ds^2 = \hat{a}^2(x)(\gamma_{\mu\nu} dx^\mu dx^\nu + dx^2),
\end{equation}
as shown in Appendix~\ref{coordinate:systems}. The behavior is
therefore different depending on the sign of $y_0$. If it is negative,
(\textit{i.e.} if $T_0<0$), then we can take
the $L\to \infty$ limit by making $T_L^\pm=1$. This in turn is
obtained for $w_L=12$, with $v_L\geq -1$ if we are in the $(T_L^+)$
branch and $v_L \leq -1$ in the $(T_L^-)$ branch. Now (\ref{Cg0}) shows
in this case that the graviton
zero mode is non-normalizable and decouples from the spectrum whereas
the radion and massive gravitons remain in the spectrum.
If $y_0$ is positive, the limit $L \to y_0$ is obtained by taking
$w_L \to -\infty$, that makes $T_L^+ \to - \infty$ or $v_L \to 0$,
making $T_L^- \to - \infty$. In that case (\ref{Cpsi}) shows that
it is the radion which is
non-normalizable while the graviton zero mode and the massive
modes remain in the spectrum. Apart from the fact that either the
graviton zero mode or the radion decouple, most of
the features of the spectrum generalize to the infinite extra
dimension case. One interesting example is the $(T_0^+,T_L^-)$ branch 
with
negative $T_0$. In that case we had regions where neither the radion
nor any of the massive gravitons was a ghost or a tachyon but the graviton
zero mode was always a ghost in such regions. Since $T_0<0$, however,
we can take the $L \to \infty$ limit and the graviton zero mode
decouples. Thus, this is another example that is tachyon and
ghost--free. Note that this is a standard solution but contrary to the
previously found ones, it has $v_L<0$.

To sum up, we have found that only the $(T_0^+,T_L^+)$ branch with
both $v_0$ and $v_L$ positive is free of ghosts and tachyons
(Fig.~\ref{TpTp:dS},
hatched area) in the case of a finite extra dimension. For an infinite
extra dimension, there are also tachyon and ghost--free regions with
$v_L<0$ in the $(T_0^+,T_L^-)$ branch.
We have seen examples
of self--accelerating solutions, $(T_0^-,T_L^+)$
with $T_0$ positive for instance, where neither the radion nor any
massive graviton is a ghost but the massless graviton is
(Fig.~\ref{TmTp:dS}, hatched area). We have seen
examples of standard solutions, $(T_0^+,T_L^+)$ for instance, where
there is a ghost in the massive graviton/radion
system (Fig.~\ref{TpTp:dS}, light unhatched area). From these
examples we conclude that the DGP ghost (the ghost present in the
self--accelerating branch of the DGP model) is not necessarily 
correlated
with the self--accelerating branch in our more general warped
backgrounds.

\section{Conclusions \label{conclusions}}

The accelerating expansion of the Universe is one of the most
exciting mysteries in modern science. The existence of self--accelerating
solutions in braneworld gravity models suggests the possibility
that an extra dimension, rather than dark energy, may explain
this phenomenon. Unfortunately the original DGP model has been criticized for
the presence of ghosts in the weakly-coupled regime of the
self--accelerating solution. 
Generically either the radion or the
longitudinal component of a massive graviton is
a ghost. 

Given the obvious importance of self--accelerating solutions,
we have investigated the presence of ghosts in a
generalization of the DGP model. Our model uses a warped $AdS_5$ bulk
and two branes with arbitrary tensions and brane-localized curvature
terms. Depending on functions of the input brane parameters which
we have denoted as $T_0$, $T_L$, 
the $AdS_5$ bulk is sliced
into sections of negative, positive, or zero curvature, giving rise to
$AdS_4$, $dS_4$, or $M_4$ branes, respectively. For each value of the
input brane parameters, there are in general two different static backgrounds
per brane; the four resulting branches are denoted by
$(T_0^{\pm}$,$T_L^{\pm})$. 
In the $dS_4$ or flat case, the $(T_0^-,T_L^+)$ and $(T_0^-,T_L^-)$ branches 
provide self--accelerating solutions.

We have performed a comprehensive analysis of the spectrum for both
$AdS_4$ and $dS_4$ branes and every branch, with special emphasis on the
presence of ghosts. The results can be summarized as follows. For the
$AdS_4$ brane on the $(T_0^+,T_L^+)$ branch, the spectrum is free of
ghosts and tachyons provided both localized curvature terms are positive;
on all other $AdS_4$ branches, for all values of the input parameters,
the spectrum has either ghosts or tachyons.
For $dS_4$ branes on the $(T_0^+,T_L^+)$ branch the spectrum is also free of
ghosts and tachyons provided both localized curvature terms are positive.
On all other $dS_4$ branches, with finite $L$, 
for all values of the input parameters,
the spectrum has either  ghosts or tachyons. This feature of our results
is discouraging.

We find, however, solutions that
depart from the general behavior of the DGP model in interesting ways.
We have
self--accelerating solutions which are free of DGP ghosts,
\textit{i.e.}, neither the radion nor any massive
graviton are ghosts; in these cases the graviton zero mode turned out to be
a ghost. 
In some of these solutions, one can take the second brane to infinity
so that the (ghost) graviton zero mode decouples, thus obtaining
another example of ghost and tachyon--free model.
We also have solutions which are not self--accelerating,
but nevertheless have a DGP ghost.

Although there is no self--accelerating
solution that is totally ghost-free, the fact that
the ghost present in the DGP model is not correlated with the
self--accelerating branch makes these models an excellent laboratory to
study self--acceleration. One can hope that simple
modifications of these models could get rid of ghosts altogether.
For example, in the case of a flat bulk it was recently suggested
that adding a Gauss-Bonnet term to the bulk action will exorcise
the DGP ghost \cite{deRham:2006pe}. Since in some cases the
ghost is a radion, one could also attempt to modify the bulk scalar
dynamics, in the style of Goldberger and Wise \cite{Goldberger:1999uk}
(see also \cite{Izumi:2006ca}).
In this context, it is important to also consider the effects of
brane matter, which requires a fully time-dependent analysis rather
than the quasi-static approach taken here.
Another promising direction is the study of exact solutions, using the
methods developed in \cite{Kaloper:2005az}.

\begin{acknowledgments}
We are grateful for discussions with G. Dvali, R. Gregory,
and N. Kaloper. MC and JL acknowledge the support of the Aspen
Center for Physics, where part of this work was completed.
Fermilab is operated by Universities Research Association 
Inc. under Contract No. DE-AC02-76CHO3000 with the U.S. Department
of Energy.
\end{acknowledgments}

\appendix

\section{Coordinate systems \label{coordinate:systems}}

In this Appendix we will give the $AdS_5/(A)dS_4$ background 
solutions in a different coordinate system that has been used in other
studies of 
brane induced gravity models and is also the preferred one in
cosmological studies. This coordinate system is 
conformal:
\begin{equation}
ds^2=\hat{a}^2(x)(\gamma_{\mu\nu} dx^\mu dx^\nu+dx^2),
\end{equation}
where
\begin{equation}
a^2(y)=\hat{a}^2(x), \quad dy^2=\hat{a}^2(x)dx^2.
\end{equation}
The coordinate change is of course different for the $dS_4$ and
$AdS_4$ cases. 
\begin{itemize}
\item $\mathbf{AdS_4}$ \textbf{case}:
\begin{equation}
\hat{a}(x)=\frac{H}{k} \frac{1}{\sin H(x-x_0)},
\end{equation}
with
\begin{equation}
x-x_0=-\frac{\mathrm{i}}{2H} \ln \left(\frac{\sinh k(y-y_0)-\mathrm{i}}
{\sinh k(y-y_0)+\mathrm{i}}\right),
\end{equation}
and
\begin{equation}
y=y_0-\frac{1}{k} 
\sinh^{-1} \big[ \cot H(x-x_0) \big].
\end{equation}
Requiring that $x=0$ for $y=0$ we obtain
\begin{equation}
x_0=-\frac{\mathrm{i}}{2H} \ln \left(\frac{\sinh ky_0 -\mathrm{i}}
{\sinh ky_0+\mathrm{i}}
\right).
\end{equation}
In particular it is easy to check that
\begin{eqnarray}
y- y_0 \to 0^\pm & \Rightarrow & x-x_0\to \mp \pi/(2H), \\
y \to \infty &\Rightarrow & x- x_0 \to 0^-.
\end{eqnarray}
\item $\mathbf{dS_4}$ \textbf{case}:
\begin{equation}
\hat{a}(x)=
\frac{H}{k} \frac{1}{\sinh H(x-x_0)},
\end{equation}
with
\begin{equation}
x-x_0=-\frac{1 
}{2H} \ln \left(\frac{\cosh k(y-y_0)-1}
{\cosh k(y-y_0)+1}\right),
\end{equation}
and
\begin{equation}
y=y_0-\frac{\tilde{\epsilon}_0}{k} 
\cosh^{-1} \big[ \coth 
H(x-x_0) \big],
\end{equation}
where $\tilde{\epsilon}_0=\mathrm{sign}~T_0$.
Again, requiring that $x=0$ for $y=0$ we obtain
\begin{equation}
x_0=\frac{1 
}{2H} \ln \left(\frac{\cosh ky_0 -1}{\cosh ky_0+1}
\right).
\end{equation}
In this case we have
\begin{eqnarray}
y\to y_0 & \Rightarrow & x \to 
\infty, \\
y \to \infty &\Rightarrow & x- x_0 \to 0.
\end{eqnarray}
\end{itemize}

\section{Detailed calculations for $dS_4$ branes 
\label{detailed:calculations}
}

We reproduce in this appendix the detailed calculations that lead to
the determination of ghosts and tachyons for $dS_4$ branes. 
For convenience, we will use
$(v, T)$ as independent variables. 
The first step is to
characterize the region of parameter space satisfying
$\bar{C}_\psi=0$. Let us fix the value of $(v_L, T_L)$, 
and solve $\bar{\mathcal C}_\psi(v_0^*,T_0^*) = 0$ for $v_0^*$:
\be\label{eqn:v0*}
v_0^* = \frac{-2T_0^*-2-\chi_L}{(T_0^*+1)^2+\chi_L T_0^*}\,,
\ee
where
\be\label{eqn:chi}
\chi_L = \frac{(T_L-1)(2-v_L+v_L T_L)}{1+v_L T_L}\,.
\ee
In the $(v_0, T_0)$--plane, 
the allowed regions are bounded by $T_0 = \pm1$ or
$T_0 = -T_L$, and $T_0=-1/v_0$. 
For example, the allowed region for the $(T_0^-, T_L^-)$ branch with
$T_0^- < 0$ lies in the fourth quadrant with 
$T_0 < -T_L$ and $T_0 < -1/v_0$. In this case, in order for the
$\bar{C}_\psi=0$ line to fit in the allowed region,  
it should be satisfied that $v_0^*$ of (\ref{eqn:v0*}) be larger than
$-1/T_0^*$.  
This inequality can change sign for some of the branches. 

Therefore, to see whether $\bar{C}_\psi=0$ line lies inside or outside of the 
allowed region, we need to check the sign of
\be\label{eqn:v+1/t}
v_0^* + \frac{1}{T_0^*} = \frac{1-T_0^*{}^2}{T_0^*} \frac{1}{T_0^*{}^2
+ (2+\chi_L)T_0^* + 1}\,, 
\ee
whose only nontrivial part is
\be
{\mathfrak S}(T_0) = T_0^*{}^2 + (2+\chi_L)T_0^* + 1\,.
\ee
Since $T_0 < -T_L < -1$ or $1 < T_0 < -T_L$, from the elementary
analysis on quadratic functions 
it would suffice to determine the signs of
\be\label{eqn:}
{\mathfrak S}(-T_L) = \frac{1-T_L^2}{1+v_L T_L}\,,\quad {\mathfrak
S}(1) = 4 + \chi_L\,, 
\ee
and the location of the symmetry axis, $l=-(2+\chi_L)/2$, relative to
$-T_L$ or $1$. Note that  
the sign of ${\mathfrak S}(-T_L)$ is the opposite of $\epsilon_L$.
The result is summarized in the following table:

\vspace{5pt}
\begin{tabular}{c|c|c c c|c|c|c}
\multicolumn{2}{c|}{} & ${\mathfrak S}(-T_L)$ & ${\mathfrak S}(1)$ &
$l$ & sign[(\ref{eqn:v+1/t})] &  
allowed region & location of $C_\psi=0$\\
\hline
           & $(T_0^+, T_L^+)$ & $-$ &  &  & $+ \to -$ as $T_0$
increases & $v_0  +1/T_0<0$ & inside or outside\\ 
\cline{2-8}                  
$T_0 < 0$ & $(T_0^+, T_L^-)$ & $+$ &  & $> -T_L$ & $+$ & 
$v_0 +1/T_0<0$ & outside \\ 
\cline{2-8} 
           & $(T_0^-, T_L^+)$ & $-$ &  &  & $+ \to -$ as $T_0$
increases & $v_0 +1/T_0 > 0$ & inside or outside\\ 
\cline{2-8}                  
           & $(T_0^-, T_L^-)$ & $+$ &  & $> -T_L$ & $+$ & $v_0 +1/T_0 >
0$ & inside  \\ 
\hline
           & $(T_0^+, T_L^+)$ & $-$ & $-$ &  & $+$ & $v_0 +1/T_0 > 0$ 
& inside \\
\cline{2-8}                   
           & $(T_0^+, T_L^-)$ & $+$ & $+$ & $< 1$ & $-$ & $v_0 +1/T_0 >
0$ & outside \\ 
$T_0 > 0$  &                  &     & or $-$ &  & $+ \to -$ as $T_0$
increases & $v_0 +1/T_0 > 0$ &inside or outside\\ 

\cline{2-8}
           & $(T_0^-, T_L^+)$ & $-$ & $-$ &  & $+$ & $v_0 +1/T_0 < 0$ &
outside \\ 
\cline{2-8}                  
           & $(T_0^-, T_L^-)$ & $+$ & $+$ & $< 1$ & $-$ & $v_0 +1/T_0<
0$ & inside \\ 
           &                  &     & or $-$ &  & $+ \to -$ as $T_0$
increases & $v_0 +1/T_0 < 0$ & inside or outside\\ 
\end{tabular}
\vspace{5pt}

\noindent A blank in the tables denotes that the corresponding
information does not affect the result. $+ \to -$ as $T_0$ increases
means that both signs are possible, depending on the parameters. 
Comparing the sign of (\ref{eqn:v+1/t}) with the allowed region for
each case (columns 4 and 5 in the table) we obtain the classification
of the last column in the table.

Let us assume the line of $\bar{\mathcal \mathcal{C}}_\psi = 0$ is inside
the allowed region, 
with $(v_0^*,T_0^*)$ representing the locus.
The treatment of the solution is different depending on the sign 
of $T_0$. 
If $T_0$ is negative, so that $z$ is positive, we get, 
from (\ref{eqn:deteq},\ref{Cpsi}), 
\bea
{\mathfrak D}(q=2)&=&4(T_0+1)(2+v_0+v_0 T_0)(1+v_L T_L) 
+ 4(T_L-1)(2-v_L+v_L T_L)(1+v_0 T_0)\nn\\
&=&-\frac{8}{3}(1+v_0 T_0)(1+v_L T_L) \bar{\mathcal C}_\psi \, ,
\label{eqn:dq2=cpfntl:app}
\eea
and therefore if $\bar{\mathcal C}_\psi = 0$ has a solution, there is
a KK graviton mode with $q=2$. 
Let us now consider how $\bar{\mathcal C}_\psi$ and ${\mathfrak D}$
change as we vary $T_0$ around the solution: 
for given $L$-parameters, when $s=(q=2, T_0=T_0^*)$ solves ${\mathfrak
D} = 0$,  
at $(q=2+\delta q, T_0 = T_0^* + \delta T_0)$ which is another
solution in a small neighborhood of $s$,  
we have
\bea\label{eqn:dexp}
0 &=& {\mathfrak D}(T_0^* + \delta T_0, 2+\delta q) = {\mathfrak D}(s) 
+ \frac{\partial {\mathfrak D}}{\partial T_0}\Big|_s \delta T_0
+ \frac{\partial {\mathfrak D}}{\partial q}\Big|_s \delta q \nn\\
&=& 0 - \frac{8}{3}(1+v_0T_0)(1+v_LT_L) \,
\frac{\partial \bar{\mathcal C}_\psi}{\partial T_0}\Big|_{T_0^*} \delta T_0
+ \frac{\partial {\mathfrak D}}{\partial q}\Big|_s \delta q \,,
\eea
\ie,
\be\label{eqn:dq/dz}
\frac{\delta q}{\delta T_0} = \frac{8}{3}\frac{(1+v_0T_0)(1+v_LT_L)}
{\partial {\mathfrak D}/\partial q|_s} \,
\frac{\partial \bar{\mathcal C}_\psi}{\partial T_0}\Big|_{T_0^*} \,.
\ee
To evaluate $\partial {\mathfrak D}/\partial q$, we use \cite{Bateman}
3.7 (6),  
\cite{gradrhyz} 4.224.9 and \cite{Bateman} 3.6.1 (8) to get
\bea
\partial_\nu P_\nu^0(z)|_{\nu=0} &=& \ln \frac{z+1}{2}\,, \\
\partial_\nu P_\nu^{-1}(z)|_{\nu=0} &=& -\sqrt{\frac{z-1}{z+1}} 
+ \sqrt{\frac{z+1}{z-1}} \ln \frac{z+1}{2} \,, \\
\partial_\nu P_\nu^{-2}(z)|_{\nu=0} &=& -\frac{3z+1}{4(z+1)} 
+ \frac{z+1}{2(z-1)} \ln \frac{z+1}{2} \,.
\eea
Also using \cite{gradrhyz} 8.712, 2.727.2 and \cite{Bateman} 3.6.1 (5)
we obtain 
\bea
\partial_\nu Q_\nu^0(z)|_{\nu=0} &=& \frac{1}{2}\ln 2 \ln\frac{z+1}{z-1}
+ \frac{1}{4}\ln(z^2-1)\ln\frac{z-1}{z+1} \nn\\
&&- \frac{1}{2}{\rm Li}_2\Big(\frac{2}{1+z} \Big)
+ \frac{1}{2}{\rm Li}_2\Big(\frac{2}{1-z} \Big) \,, \\
\partial_\nu Q_\nu^1(z)|_{\nu=0} &=& 
\frac{1}{2\sqrt{z^2-1}}\Big( z \ln\frac{z+1}{z-1} + \ln\frac{z^2-1}{4}
\Big) \,,\\ 
\partial_\nu Q_\nu^2(z)|_{\nu=0} &=& 
-\frac{1}{2(z^2-1)} \Big\{ (z^2+1)\ln\frac{z+1}{z-1} +
2z\ln\frac{z^2-1}{4} \Big\} \,, 
\eea
where ${\rm Li}_2(z)$ is the dilogarithm function, defined as
\be
{\rm Li}_2(z) = \sum_{k=1}^\infty \frac{z^k}{k^2} 
= \int_z^0 \frac{\ln (1-t)}{t} \d t \,.
\ee
Then, $\partial {\mathfrak D}/\partial q|_{q=2}$ becomes
\bea
&&\hspace{-30pt} \frac{\partial {\mathfrak D}}{\partial q}\Big|_{q=2} =
4(1+v_L T_L) \Big\{ 2(T_0+1)(1+v_0 T_0) 
+ (2-v_0 + v_0 T_0)(-T_0+1) \ln\frac{-T_0+1}{2}\Big\} \nn\\
&&\hspace{-10pt} +(2-v_L + v_L T_L)(T_L-1)
\Big\{ (v_0 + 2T_0 + v_0 T_0^2)\ln\frac{-T_0-1}{-T_0+1} 
+ 2(1+v_0T_0)\Big(1+\ln\frac{T_0^2-1}{4}\Big) \Big\} \nn\\
&&\hspace{-10pt} + 4(1+v_0 T_0) \Big\{ 2(T_L-1)(1+v_L T_L) 
- (2+v_L + v_L T_L)(T_L+1) \ln\frac{T_L+1}{2}\Big\} \nn\\
&&\hspace{-10pt} -(2+v_0 + v_0 T_0)(T_0+1)
\Big\{ (v_L + 2T_L + v_L T_L^2)\ln\frac{T_L-1}{T_L+1} 
- 2(1+v_LT_L)\Big(1+\ln\frac{T_L^2-1}{4}\Big) \Big\} \,. \nn\\
\eea
At $T_0=T_0^*\,$, (\ref{eqn:dq2=cpfntl:app})$=0$ simplifies the above into
\bea\label{eqn:dDats2}
\frac{\partial {\mathfrak D}}{\partial q}\Big|_s 
&=& \frac{(T_0^*{}^2-1)(T_L^2-1)}{(T_L-1)^2-\chi_L T_L} \cdot 
\frac{8(T_0^*+T_L)-\chi_L^2 \ln \frac{-T_0^*-1}{T_L-1}
+(\chi_L+4)^2 \ln\frac{-T_0^*+1}{T_L+1} }{(T_0^*+1)^2+\chi_L T_0^*}\,,
\eea
and (\ref{eqn:dq/dz}) becomes
\be\label{eqn:dq/dzfin}
\frac{\delta q}{\delta T_0}\Big|_s = \frac{8}{3 B(T_0;T_L,\chi_L)} 
\frac{\partial \bar{\mathcal C}_\psi}{\partial T_0}\Big|_{T_0^*} \,,
\ee
where
\be\label{eqn:b}
B = 8(T_0^\ast+T_L)-\chi_L^2 \ln \frac{-T_0^\ast-1}{T_L-1}+(\chi_L+4)^2
\ln\frac{-T_0^\ast+1}{T_L+1}\,. 
\ee
$B$ is a growing function of $T_0$,
\be
\frac{\d B}{\d T_0} = 2\frac{(2+2T_0+\chi_L)^2}{T_0^2-1} > 0\,.
\ee
For $L>0$ we have $T_0 < -T_L$ and therefore 
$B$ is bounded above by $B(-T_L) = 0$, or in other
words, it is always negative. If $L<0$, on the other hand, we have
$T_0>-T_L$ and therefore $B$ is  bounded below by $0$, so it is positive.  
Thus, when $\bar{\mathcal C}_\psi = 0$ has a solution in the physical
region (with $L>0$), 
the slope of the mass of the first KK graviton mode when it passes $q=2$ 
and that of $\bar{\mathcal C}_\psi$ when it crosses zero have opposite signs, 
and therefore when the first KK graviton mode is heavier than $2H^2$ the
radion is a ghost and \textit{vice versa}.  
If $\bar{\mathcal C}_\psi = 0$ has a solution outside the physical
region (where $L<0$), then both the radion and the longitudinal
component of a massive graviton are ghosts at one side of the line (in
the $(v_0,w_0)$ plane) and neither of them is at the other side.

If $T_0$ is positive, then $z$ is negative in the physical region and
we have to use the wave functions evaluated at $-z$ for the massive
gravitons.
This introduces some extra minus signs, so that 
now we have
\bea\label{eqn:dq2=cpfntl-}
{\mathfrak D}(q=2) &=& \frac{8}{3}(1+v_0 T_0)(1+v_L T_L) \bar{\mathcal
C}_\psi \,, \\ 
\label{eqn:dq/dz-}
\frac{\delta q}{\delta T_0}\Big|_s 
&=& -\frac{8}{3}\frac{(1+v_0T_0)(1+v_LT_L)}{\partial {\mathfrak
D}/\partial q|_s} \, 
\frac{\partial \bar{\mathcal C}_\psi}{\partial T_0}\Big|_{T_0^*} \,,\\
\label{eqn:dDats2-}
\frac{\partial {\mathfrak D}}{\partial q}\Big|_s 
&=& -\frac{(T_0^*{}^2-1)(T_L^2-1)}{(T_L-1)^2-\chi_L T_L} \cdot 
\frac{8(T_0^*+T_L)-\chi_L^2 \ln \frac{T_0^*+1}{-T_L+1}
+(\chi_L+4)^2 \ln\frac{T_0^*-1}{-T_L-1} }{(T_0^*+1)^2+\chi_L T_0^*}\,.
\eea
These extra minuses get cancelled among themselves, and we still get
the same final result,  
(\ref{eqn:dq/dzfin}-\ref{eqn:b}). Thus independently of the sign of
$T_0$, it is always true that, near the locus of points satisfying
$\bar{\mathcal{C}}_\psi=0$, 
either the radion or the longitudinal component of a
massive graviton with $m^2 < 2 H^2$ is a ghost if parameters are
chosen from inside the physical
region and neither of them is to one side of the region
$\bar{\mathcal{C}}_\psi=0$ if that region has $L<0$.

Let us now compute the limit $\mathfrak{D}(q\to -\infty)$, which is
equivalent to $\nu \to \infty$, where $\nu$ is the index of the
corresponding Legendre function. The calculation is again different for
the two signs of $T_0$. We start considering the case of negative
$T_0$, so that $z$ is positive. 
Using 8.723 1 from \cite{gradrhyz} we find 
\begin{equation}
P^{-2}_{\nu\to \infty}(\cosh \alpha) \sim \frac{1}{\sqrt{\pi}} \frac{
e^{(\nu+1)\alpha}} {\sqrt{e^{2\alpha}-1}} \nu^{-5/2},
\end{equation}
where we have defined $z\equiv \cosh \alpha$. Then
\begin{equation}
(1-z^2) \dbyd{z} P^{-2}_{\nu \to \infty}(\cosh \alpha)
\sim -\frac{1}{2\sqrt{\pi}} \sqrt{e^{2\alpha}-1} e^{\nu \alpha}
\nu^{-3/2}.
\end{equation}
Using now 8.723 2 and 8.732 1 from \cite{gradrhyz},
respectively, we get
\begin{eqnarray}
Q^2_{\nu \to \infty}(\cosh \alpha) &\sim& 
\sqrt{\pi} \frac{e^{-\nu \alpha}}{\sqrt{e^{2\alpha}-1}}\nu^{3/2}, \\
(1-z^2)\dbyd{z}Q^2_{\nu \to \infty}(\cosh \alpha) &\sim& 
\frac{\sqrt{\pi}}{2} 
\sqrt{e^{2\alpha}-1} e^{-(\nu+1) \alpha} \nu^{5/2}.
\end{eqnarray}
These expressions allow us to compute the limit of the different
coefficients, that turns out to be identical for both branes
\begin{eqnarray}
a_i(q\to -\infty) &\sim& \frac{ v_i (T_i^2-1)}{\sqrt{\pi}}
\frac{e^{(\nu-1)\alpha_i}}{\sqrt{e^{2\alpha_i}-1}} \nu^{-1/2}, \\
b_i(q\to -\infty) &\sim& \sqrt{\pi} v_i (T_i^2-1)
\frac{e^{-\nu\alpha_i}}{\sqrt{e^{2\alpha_i}-1}} \nu^{7/2}.
\end{eqnarray}
We can now compute the limit of the full determinant,
\begin{eqnarray}
a_0 b_L -a_L b_0 &\sim&
v_0 v_L \frac{T_0^2-1}{\sqrt{e^{2\alpha_0}-1}} 
\frac{T_L^2-1}{\sqrt{e^{2\alpha_L}-1}}  \nu^3
\Big[
e^{-\alpha_0} e^{\nu(\alpha_0-\alpha_L)} 
-e^{-\alpha_L} e^{\nu(\alpha_L-\alpha_0)} 
\Big] \nonumber \\
&\sim&
v_0 v_L \frac{T_0^2-1}{\sqrt{e^{2\alpha_0}-1}} 
\frac{T_L^2-1}{\sqrt{e^{2\alpha_L}-1}}  \nu^3
e^{-\alpha_0} e^{\nu(\alpha_0-\alpha_L)},
\end{eqnarray}
where we have used that, in the physical region with negative $T_0$, we
have $0<T_L< -T_0 \Rightarrow 
\cosh \alpha_L < \cosh \alpha_0 \Rightarrow \alpha_L < \alpha_0$.
Thus,
\begin{equation}
\mathrm{sign}\big[\mathfrak{D}(q\to -\infty) \big]
= - \mathrm{sign}\big[v_0 v_L \big]
=  \mathrm{sign}\big[T_0 v_0 v_L \big].
\end{equation}
We can now check what happens in the case that $T_0>0$ so that
$z<0$. As we said above, 
the valid solution is then the corresponding associated
Legendre functions evaluated at $-z$. The expression of the $a_i$ and
$b_i$ is then identical to (\ref{eqn:deteq}) with the
change $\theta_i \to - \theta_i$, \textit{i.e.}
\begin{equation}\
c_0^{(T_0>0)}(T_0)=c_L^{(T_0<0)}(T_0),
\quad c_L^{(T_0>0)}(T_L)=c_0^{(T_0<0)}(T_L),
\end{equation}
where $c_i$ stands for any of $a_i$ or $b_i$. 
Thus, noting that now $
0<T_0 < -T_L \Rightarrow 
\cosh \alpha_0 < \cosh \alpha_L \Rightarrow \alpha_0 < \alpha_L$ in
the physical region, we get
\begin{eqnarray}
a_0 b_L -a_L b_0\Big|_{T_0>0} &\sim&
v_0 v_L \frac{T_0^2-1}{\sqrt{e^{2\alpha_0}-1}} 
\frac{T_L^2-1}{\sqrt{e^{2\alpha_L}-1}}  \nu^3
\Big[
e^{-\alpha_0} e^{\nu(\alpha_0-\alpha_L)} 
-e^{-\alpha_L} e^{\nu(\alpha_L-\alpha_0)} 
\Big] \nonumber \\
&\sim&
-v_0 v_L \frac{T_0^2-1}{\sqrt{e^{2\alpha_0}-1}} 
\frac{T_L^2-1}{\sqrt{e^{2\alpha_L}-1}}  \nu^3
e^{-\alpha_L} e^{\nu(\alpha_L-\alpha_0)},
\end{eqnarray}
and therefore
\begin{equation}
\mathrm{sign}\big[\mathfrak{D}^{(T_0>0)}
(q\to -\infty) \big]
= \mathrm{sign}\big[v_0 v_L \big]
= \mathrm{sign}\big[T_0v_0 v_L \big].
\end{equation}

\section{Flat bulk limit}

In this appendix we will show how to obtain 
the limit of a flat bulk ($k \to
0$) for $dS_4$ branes.
Taking the $k\to 0$ limit in the 
$\rm AdS_5/dS_4$ background needs special care
because our usual reparametrization of  
the extra dimensional coordinate $y$, $z=\coth k(y-y_0)$, breaks down
and the input parameters, $v_i=k M_i^2/M^3$ and $w_i=V_i/2kM^3$, are
ill-defined. Therefore, 
it is easier to 
redo our analysis directly with $k=0$. That is, we
work on an $\rm M_5/dS_4$ background. Let us define the following
combinations of brane parameters and the fundamental Planck mass,
\begin{equation}
\lambda_i \equiv \frac{M_i^2}{M^3}, \quad
U_i \equiv \frac{V_i}{2M^3}.
\end{equation}

Solving for the background is straightforward, and we get
\be
G^{(0)}_{MN} \d x^M \d x^N = a^2(y) \gamma_{\mu\nu} \d x^\mu \d x^\nu
+ \d y^2\,, 
\ee
where $a= 1+\epsilon_0 {\mathcal H} y$ with $\epsilon_0=+1$ for the
self--accelerating branch and $\epsilon_0=-1$ for the standard one. 
Here $\gamma_{\mu\nu}$  
is the $\rm dS_4$ metric. The values of $\mathcal H$ and $L$ are
determined by the brane-boundary equations: 
\bea
0&=&\frac{\lambda_0}{2}{\mathcal H}^2 - \epsilon_0 {\mathcal H} -
\frac{U_0}{12}\,,\\ 
0&=&\frac{\lambda_L}{2}\Big(L+\frac{\epsilon_0}{{\mathcal H}}\Big)^{-2} 
+ \Big(L+\frac{\epsilon_0}{{\mathcal H}}\Big)^{-1} - \frac{U_L}{12}\,.
\eea

We first look at the radion. Following footsteps of \S5 of
\cite{Carena:2005gq}, 
we can get the equations of motion for the massless scalar degrees
of freedom. Solving them, we obtain
\bea\label{eqn:k=0phi1}
\varphi_1(x,y) &=& -\frac{f_2(x)}{{\mathcal H}^2} +
\frac{C(x)}{a^{(3-\epsilon_0)/2}} 
+ \frac{D(x)}{a^{(3+\epsilon_0)/2}} - {\mathfrak F}(y) \psi(x) \,,\\
\varphi_2(x,y) &=& f_2(x) + \Big( \frac{a'}{a}{\mathcal F}(y) +
{\mathcal H}^2 {\mathfrak F}(y) \Big) \psi(x)\,. 
\eea
$f_2(x)$ turns out to be a pure gauge field and will be
killed. Plugging (\ref{eqn:k=0phi1}) into  
the brane-boundary equations gives
\be
\alpha_i C + \beta_i D = \frac{{\mathcal F}(y_i)}{{\mathcal H}} \psi\,,
\ee 
where
\bea
\alpha_i &=& -\frac{\theta_i (3\epsilon_0 - 1)/2 \cdot a_i^{-(1-\epsilon_0)/2} 
+ \lambda_i {\mathcal H} a_i^{-(3-\epsilon_0)/2}}
{\theta_i + \epsilon_0 \lambda_i {\mathcal H} /a_i}\,,\\
\beta_i &=& - \frac{\theta_i (3\epsilon_0 + 1)/2 \cdot a_i^{-(1+\epsilon_0)/2} 
+ \lambda_i {\mathcal H} a_i^{-(3+\epsilon_0)/2}}
{\theta_i + \epsilon_0 \lambda_i {\mathcal H} / a_i} \,,
\eea
with $a_i = a(y_i)$. 

At the massive level, a solution of the bulk equations of motion is now
\be
b_{\mu\nu}(x,y) = a^{(1-\sqrt{9-4q})/2} A_{\mu\nu}(x) +
a^{(1+\sqrt{9-4q})/2} B_{\mu\nu}(x)\,, 
\ee
and the brane-boundary equations provide the determinant equation,
which gives the mass spectrum; 
\bea\label{eqn:m5ds4deteq}
\hspace{-10pt} 0&=&\{ (3+\sqrt{9-4q}) \epsilon_0 - {\mathcal H} \lambda_0 q \}
\cdot a_L^{(-3+\sqrt{9-4q})/2} \{ (-3+\sqrt{9-4q}) (\epsilon_0 +
{\mathcal H} L) - {\mathcal H} \lambda_L q \} \nn\\ 
\hspace{-10pt} &-& \{ (-3+\sqrt{9-4q}) \epsilon_0 + {\mathcal H} \lambda_0 q \}
\cdot a_L^{-(3+\sqrt{9-4q})/2} \{ (3+\sqrt{9-4q}) (\epsilon_0 +
{\mathcal H} L) + {\mathcal H} \lambda_L q \} \,. 
\eea

Let's look into the self--accelerating branch($\epsilon_0 = +1$). 
Here, since $\alpha_0 = \alpha_L = -1$, we can fix the 
gauge function $F(y)$ such that ${\mathcal F}(y_i) = \chi \beta_i$
with the constant $\chi$ a residual 
gauge parameter. This choice kills the $C$ mode and 
\be
D = \frac{\chi}{{\mathcal H}}\psi\,,
\ee
which redefines (\ref{eqn:y1y2}) for $\rm M_5/dS_4$ background;
\be
{\mathcal Y}_1(y) = \frac{\chi}{{\mathcal H}}a^{-2}-{\mathfrak F}\,,\;\;
{\mathcal Y}_2(y) = \chi {\mathcal H} a^{-2} + \frac{{\mathcal
H}}{a}{\mathcal F}\,. 
\ee
Working out the quadratic action, we finally get 
\be\label{eqn:m5ds4cpsisa}
C_\psi = \frac{3\chi^2H}{2}\sum_i \Big( \frac{y_i}{y_i+1/H} -
\frac{1}{\theta_i+H(y_i + \lambda_i)} \Big)\,. 
\ee
Also with $\epsilon_0 = +1$, (\ref{eqn:m5ds4deteq}) becomes
\bea\label{eqn:m5ds4deteqsa}
\hspace{-10pt} 0&=&\{ (3+\sqrt{9-4q}) - {\mathcal H} \lambda_0 q \}
\cdot a_L^{(-3+\sqrt{9-4q})/2} \{ (-3+\sqrt{9-4q}) (1 + {\mathcal H}
L) - {\mathcal H} \lambda_L q \} \nn\\ 
\hspace{-10pt} &-& \{ (-3+\sqrt{9-4q}) + {\mathcal H} \lambda_0 q \}
\cdot a_L^{-(3+\sqrt{9-4q})/2} \{ (3+\sqrt{9-4q}) (1 + {\mathcal H} L)
+ {\mathcal H} \lambda_L q \} \,. 
\eea

We can match the above with the recent results of Izumi {\it et
al}. for  
braneworld models on a $\rm M_5/dS_4$ background with a finite extra
dimension \cite{Izumi:2006ca}.  
The conversion from our setup to theirs can be achieved by the
following identifications: 
$\lambda_0, \lambda_L \to 2r_c$, $V_i \to \tau_i$. Then correspondence among 
derived quantities follows: ${\mathcal H} \to H_+ = \hat H_+ + 1/r_c$, 
$L \to 1/H_- - 1/H_+$, $q \to m_i^2$. Upon making these substitutions,
(\ref{eqn:m5ds4cpsisa})  
and (\ref{eqn:m5ds4deteqsa}) turn into
\bea
C_\psi &=& 3\chi^2 (\hat H_+-H_-) \frac{1+(1+2\hat H_+ r_c)(1+2H_-
r_c)}{(1+2\hat H_+ r_c)(1+2H_- r_c)} \nn\\ 
&=& 3\chi^2 \Big\{ \frac{\hat H_+(1+\hat H_+ r_c)}{1+2\hat H_+ r_c} 
- \frac{H_-(1+H_- r_c)}{1+2H_- r_c} \Big\} \; .
\eea
This expression is a positive constant times times the expression
in eqn. (2.42) of \cite{Izumi:2006ca}.
Similarly:
\bea
\label{eqn:tanakamatchingdeteq}
0 &=& \Big(3-2H_+ r_c m_i^2 +\sqrt{9-4m_i^2} \Big) \nn\\
&&\times \Big( \frac{H_+}{H_-} \Big)^{(-1+\sqrt{9-4m_i^2})/2}
 \Big(-3 -2H_-r_cm_i^2+\sqrt{9-4m_i^2} \Big) \nn\\
&&- \Big(3-2H_+ r_c m_i^2 -\sqrt{9-4m_i^2} \Big) \nn\\
&&\quad \times \Big( \frac{H_+}{H_-} \Big)^{-(1+\sqrt{9-4m_i^2})/2}
 \Big(-3 -2H_-r_cm_i^2-\sqrt{9-4m_i^2} \Big) \; ,
\eea
which is $4H_-$ times the determinant of the expression in
eqn. (2.35) of \cite{Izumi:2006ca}.
The normal branch solutions 
($\epsilon_0 = -1$) can be worked out similarly.

\end{document}